\documentclass{article}
\usepackage[T1]{fontenc}
\usepackage[margin=1in]{geometry}
\usepackage{amsfonts,amsmath,amsthm,amssymb,mathtools}

\mathtoolsset{centercolon}
\usepackage{xfrac,nicefrac}
\usepackage{mathdots}
\usepackage{mleftright}
\let\left\mleft
\let\right\mright
\usepackage{todonotes}

\usepackage{xspace}
\xspaceaddexceptions{]\}}
\usepackage{regexpatch}

\usepackage{bm,bbm,dsfont}
\usepackage{caption}
\usepackage[normalem]{ulem}
\usepackage{enumitem}
\usepackage{needspace}

\usepackage{graphicx}
\usepackage{float}
\usepackage{subcaption}
\usepackage{tcolorbox}
\usepackage{tikz}
\usetikzlibrary{decorations.pathreplacing}
\usetikzlibrary{calc}
\usetikzlibrary{positioning}
\usetikzlibrary{arrows.meta}

\usepackage[linesnumbered,boxed,ruled,vlined]{algorithm2e}

\SetCommentSty{mycommfont}

\usepackage{thmtools,thm-restate}
\usepackage[colorlinks,citecolor=blue,linkcolor=blue,urlcolor=red]{hyperref}

\theoremstyle{plain}

\newtheorem{theorem}{Theorem}[section]
\newtheorem{lemma}[theorem]{Lemma}
\newtheorem{fact}[theorem]{Fact}
\newtheorem{observation}[theorem]{Observation}
\newtheorem{proposition}[theorem]{Proposition}

\newtheorem{claim}[theorem]{Claim}

\theoremstyle{definition}
\newtheorem{definition}[theorem]{Definition}

\makeatletter
\@for\thm@name:={theorem,lemma,fact,observation,proposition,corollary,cor,claim,conjecture,definition,remark,property}\do{%
  \expandafter\def\csname theH\thm@name\endcsname{\thesection.\arabic{theorem}}}
\makeatother

\usepackage[capitalise]{cleveref}
\crefname{algocf}{Algorithm}{Algorithms}
\Crefname{algocf}{Algorithm}{Algorithms}
\crefname{claim}{Claim}{Claims}
\Crefname{claim}{Claim}{Claims}
\crefname{fact}{Fact}{Facts}
\Crefname{fact}{Fact}{Facts}
\crefname{observation}{Observation}{Observations}
\Crefname{observation}{Observation}{Observations}

\usepackage{xurl}

\newfloat{Distribution}{htbp}{loa}
\crefname{Distribution}{Distribution}{Distributions}
\Crefname{Distribution}{Distribution}{Distributions}
\newfloat{Protocol}{htbp}{loa}
\crefname{Protocol}{Protocol}{Protocols}
\Crefname{Protocol}{Protocol}{Protocols}

\SetKwProg{Function}{Function}{:}{}

\SetKwProg{CodeBlock}{}{}{}
\SetKwProg{Repeat}{Repeat}{:}{}

\DeclarePairedDelimiter{\floor}{\lfloor}{\rfloor}

\DeclarePairedDelimiter{\bk}{(}{)}
\DeclarePairedDelimiter{\Bk}{[}{]}

\DeclarePairedDelimiterX\mysetbase[2]{\lbrace}{\rbrace}{#1\,\delimsize\vert\,#2}
\NewDocumentCommand{\myset}{sO{}m m}{%
  \IfBooleanTF{#1}%
    {\mysetbase*{#3}{#4}}%
    {\mysetbase[#2]{#3}{#4}}%
}

\DeclareMathOperator*{\E}{\mathbb{E}}

\let\Pr\PrAux
\DeclareMathOperator{\poly}{poly}

\DeclareMathOperator*{\ind}{\mathbbm{1}}

\newcommand{\eps}{\varepsilon}

\renewcommand{\emptyset}{\varnothing}
\renewcommand{\epsilon}{\eps}

\newcommand{\defn}[1]{\emph{\boldmath\textbf{#1}}}

\usepackage{regexpatch}
\makeatletter
\xpatchcmd\thmt@restatable{%
\csname #2\@xa\endcsname\ifx\@nx#1\@nx\else[{#1}]\fi
}{%
\ifthmt@thisistheone
\csname #2\@xa\endcsname\ifx\@nx#1\@nx\else[{#1}]\fi
\else
\csname #2\@xa\endcsname[{Restated}]
\fi}{}{}
\makeatother

\title{Tight Lower Bounds for Value-Dynamic Retrieval}
\author{
  Yang S. Hu\thanks{Carnegie Mellon University. \texttt{yanghu2@andrew.cmu.edu}}
  \and
  William Kuszmaul\thanks{Carnegie Mellon University. \texttt{kuszmaul@cmu.edu}}
  \and
  Jingxun Liang\thanks{Carnegie Mellon University. \texttt{jingxunl@andrew.cmu.edu}}
  \and
  Renfei Zhou\thanks{Carnegie Mellon University. \texttt{renfeiz@andrew.cmu.edu}}
}
\date{}

\begin{document}

\maketitle

\begin{abstract}
In the value-dynamic retrieval problem, a data structure stores a function $f : S \to [2^b]$ on a fixed set $S$ of $n$ keys from a universe $[U]$, and must support queries and updates to $f$ (without being required to give meaningful answers for keys outside of $S$). The classical solution, which stores the values in an array indexed by a minimal perfect hash function, uses $nb + n \log e + o(n)$ bits. We prove a matching lower bound, showing that for $b = \omega(1)$ and $U = \omega(n)$, a space bound of $nb + n \log e - o(n)$ bits is necessary. We also consider a partially-dynamic version of the problem in which only $n/k$ of the keys have dynamic values. Intuitively, one might expect the space bound in this case to drop to something like $nb + \Theta(n/k)$ bits, averaging to $O(1)$ bits of redundancy per dynamic value. Our second result says that this is not the case; instead, for $k = \omega(1)$ and $2n \le U \le 2^{\poly(n)}$, the optimal space bound becomes $nb + \Theta(n \log k / k)$, meaning that any solution must pay the asymptotic cost of encoding which $n/k$ keys (out of the $n$) have dynamic values. All of our lower bounds are information-theoretic, and are proven via one-way communication arguments.
\end{abstract}

\section{Introduction}

One of the most basic tasks that a database may be asked to solve is to store a set of values, possibly changing over time, for some fixed set of keys. Examples include the prices associated with a set of stock tickers, the permission bits associated with a set of users, the availability bits associated with a set of resources, etc. If each value is $b$ bits, and there are $n$ key-value pairs, then such a data structure must necessarily use $nb$ bits of space. Perhaps surprisingly, the question of how close a data structure can come to meeting this space bound (i.e., how much redundancy a data structure must incur) has remained open. This is the question that we study in the current paper.

Formally, in the \emph{value-dynamic retrieval problem}, a data structure is given a set $S \subseteq [U]$ of $n$ keys and is responsible for storing a dynamically changing function $f:S \rightarrow [2^b]$ that maps the keys to $b$-bit values. The data structure must support two operations:
    \begin{itemize}
        \item \textsc{Update}: Given $x\in S$ and $v\in [2^b]$, update $f(x)$ to $v$.
        \item \textsc{Query}: Given $x\in S$, return $f(x)$.
    \end{itemize}
Note that both operations require the \emph{user} to guarantee that $x \in S$. Should the user invoke an \textsc{Update} operation on some key $x \not\in S$, the data structure is allowed to exhibit arbitrary (otherwise incorrect) behavior in the future; and, should the user invoke \textsc{Query} on some key $x \not\in S$, the data structure may return any value $v \in [2^b]$. This distinction turns out to be important, as it allows for solutions that are much more space efficient than would otherwise be possible.

The classical solution to this problem is to store a \emph{minimal perfect hash function} $h$ \cite{fredman1984storing,mehlhorn1982program,fredman1984size,HT01} mapping each key $x \in S$ to a distinct position in some array $A$ of size $n$. The data structure can then store the value $f(x)$ directly in position $A[h(x)]$. The hash function $h$ can be constructed to use $n \log e + o(n)$ bits, while supporting $O(1)$-time operations \cite{HT01}. Since the array $A$ itself is only $nb$ bits, the total space usage is
\begin{equation}n \log e + nb + o(n)
\label{eq:mphbound}
\end{equation}
bits. This bound is often much better than the $\Theta(n \log (U/n) + nb)$ bits that would be needed to store the keys/values explicitly in, say, a hash table. Especially for applications with relatively small values, this can be the difference between a data set fitting entirely in cache (or in memory) versus overflowing out of cache (or onto disk).

In general, any solution must use space $nb$ bits to store the $n$ values. A solution that uses $nb + R(n, b)$ bits is said to have \emph{redundancy} $R(n, b)$. The classical solution achieves redundancy $n \log e + o(n)$. The main open question in the area is whether this bound is optimal.

It has been known since the early 1980s \cite{mehlhorn1982program,fredman1984size} that any minimal perfect hash function requires $n \log e - O(\log n)$ bits, so the redundancy in \eqref{eq:mphbound} is necessary for any solution that uses minimal perfect hashing. What is less clear is whether alternative solutions may be able to do better. Kuszmaul and Walzer \cite{kuszmaul2024space} prove that any solution must incur redundancy $\Omega(n)$ (concretely, at least $0.16n - o(n)$ bits in general, and at least $0.29n - o(n)$ bits when $b = \omega(1)$). They also observe that, for $b = O(1)$, it is possible to achieve redundancy $n \log e - \Omega(n)$. For $b = \omega(1)$, however, the classical minimal-perfect-hashing solution has remained the best known result. It is conjectured that, in this regime, the space bound \eqref{eq:mphbound} achieved by minimal perfect hashing should be optimal \cite{kuszmaul2024space,kuszmaul2025fingerprint}.

\paragraph{Minimal perfect hashing is optimal.}
In this paper, we prove that, indeed, when $b = \omega(1)$ and $U = \omega(n)$, any solution must use space at least
$$n \log e + nb - o(n)$$
bits (\cref{thm:super-constant-v}).\footnote{Some assumption on the size of the universe is necessary: if $U = n$, then $S = [U]$, and the values can simply be stored in an array of $nb$ bits.} Thus, for $b = \omega(1)$, the classical minimal-perfect-hashing solution is optimal up to additive $o(n)$ terms. The lower bound applies regardless of time efficiency, and is purely information-theoretic.

Another way to interpret this result is in terms of the \emph{cost of dynamism}. Static retrieval (in which the function $f$ never changes) can be solved using $nb + o(n)$ bits \cite{dietzfelbinger2008succinct,porat2009optimal,dietzfelbinger2019constanttime}. Kuszmaul and Walzer \cite{kuszmaul2024space} showed that the cost of making the values dynamic is $\Theta(n)$ bits. Our result shows that, for $b = \omega(1)$, this cost is more precisely $n \log e \pm o(n)$, the cost of a minimal perfect hash function.

For $b=O(1)$ (i.e., $b$ does not grow with $n$), we know from \cite{kuszmaul2024space} that the redundancy can be reduced to $n \log e - \Omega(n / 2^b)$. We show that, in the other direction, when $U = \omega(n)$, the redundancy must be at least $n \log e - O(n \sqrt{(\log b) / b}) - o(n)$ (\cref{thm:constant-v}). So, as $b$ grows, the optimal redundancy converges to $n \log e$ from below, although the exact rate of convergence remains open.

\paragraph{High-level approach.}
The lower bound is achieved by a communication-complexity argument. We assume for the sake of contradiction that there exists a solution that saves $\Omega(n)$ bits of space, and then we show how to leverage this in order to obtain an impossibly good one-way communication protocol.

The basic idea is to ask, for each key $x \not\in S$, whether updating $x$ would corrupt the value of any key in $S$. If there is some key $x \not\in S$ that does \emph{not} interfere with any key in $S$, then the data structure is secretly capable of storing an $(n + 1)$-th value, and we can use this to encode extra information. On the other extreme, if every key outside of $S$ interferes with many keys in $S$, then one can argue that the data structure must implicitly store a large amount of information about $S$. Intuitively, in this case, the set $S$ behaves very \emph{unlike} a random set of size $n$, because most sets of size $n$ have a large amount of internal interference. More concretely, if we reveal $S$ key by key, then for any partial reveal of $S$, we can deduce that the rest of $S$ must come only from keys that \emph{do not} interfere with those revealed so far; and if the number of such keys is small, then the data structure must implicitly encode information about $S$.

With this high-level idea in mind, we structure the proof in two parts. First, as a warm-up (\cref{sec:warm-up}), we give a specialized argument for the case where the values are \emph{very} large ($b = \omega(n)$). What makes this case simpler is that, even if the data structure can implicitly encode a \emph{single} extra value, that is already enough to force a redundancy greater than $n \log e$ bits. Then, we show how to adapt the construction to work for smaller values ($b = \omega(1)$) (\cref{sec:super-constant-v}). Here, it is no longer enough to find a single non-interfering key, so we must instead reason about large \emph{sets} of keys and introduce a more subtle notion of what it means for such a set to interfere with $S$. Finding the right definition of such a set, so that the protocol saves $n \log e - o(n)$ bits regardless of whether such a set exists, is the most technically interesting step in the proof.

We remark that the final proof is, in some sense, surprisingly simple when compared to similarly sharp space lower bounds that have been proven for other problems (e.g., for dynamic filters \cite{kuszmaul2025fingerprint} or dynamic retrieval \cite{kuszmaul2025tight}). Anecdotally, the lower bound is the simplest example of a proof that the authors have encountered, where GPT 6 Astra Pro does not appear to be able to derive the solution using the ``keep trying'' method (in which it is told repeatedly to keep trying).

\paragraph{Partially value-dynamic retrieval.}
In many applications, only some of the keys need to have their values updated. This motivates the following intermediate problem between static and value-dynamic retrieval. The data structure is given $n$ keys, of which only $n/k$ (designated at construction time) are \emph{dynamic}, meaning that their values may be updated, and the remaining keys are \emph{static}. If $k = 1$, this is just value-dynamic retrieval, and as $k \rightarrow \infty$, the problem approaches static retrieval.

At first glance, it would seem natural that the optimal redundancy for this problem would be something like $(n / k) \log e$ bits, i.e., $\log e$ bits for each dynamic key. We show that, perhaps surprisingly, this is not the case. For $k = \omega(1)$ and $U \ge 2n$, any solution must use space
$$nb + \Omega\left(\frac{n}{k} \log k\right)$$
bits, and this holds regardless of the value size $b$ (\cref{thm:partially-dynamic}). We also give a simple proof in \cref{sec:partially-dynamic-upper}, that this bound is tight up to constant factors in the redundancy when $U \le 2^{\poly(n)}$. Thus, for the partially-dynamic version of the problem, there is a fundamental cost of $\Theta(\log k)$ bits per dynamic key.

Intuitively, this cost comes from the fact that the data structure must, in some sense, remember \emph{which} $n/k$ keys (of the $n$ total) are dynamic, which is $\log \binom{n}{n/k} = \Theta((n/k) \log k)$ bits of information. The proof reuses the ideas behind our main lower bound, with the static keys playing the role of the keys that must be preserved, and the set of dynamic keys playing the role of the information to be compressed. We show that, if the data structure does not reveal which keys are dynamic, then either the dynamic keys can be compressed given the static keys, or it must have unreasonably many distinct memory configurations that agree on the values of the \emph{static} keys; and we then argue that these configurations can be used to encode enough extra information to imply the lower bound.

\paragraph{Paper outline.}
We begin in \cref{sec:prelim} by formally defining value-dynamic retrieval and partially value-dynamic retrieval, along with the basic facts about encoding arguments that all of our lower bounds use. More specialized tools are introduced in the sections where they are first needed.

\Cref{sec:warm-up,sec:super-constant-v,sec:constant-v} concern value-dynamic retrieval. In \cref{sec:warm-up}, as a warm-up, we prove that any value-dynamic retrieval data structure must use $nb + n \log e - o(n)$ bits when the values are very large ($b = \omega(n)$). In \cref{sec:super-constant-v}, we extend this lower bound to all $b = \omega(1)$. Then, in \cref{sec:constant-v}, we show how to extend the lower bound to obtain an explicit dependence on $b$, so that the overall bound is $nb + n \log e - O(n \sqrt{(\log b)/b}) - o(n)$ bits.

\Cref{sec:partially-dynamic} concerns partially value-dynamic retrieval: we prove our lower bound of $nb + \Omega((n/k) \log k)$ bits, for all value sizes $b$. Finally, in \cref{sec:partially-dynamic-upper}, we give a matching upper bound of $nb + O((n/k) \log k)$ bits for partially value-dynamic retrieval when $U \le 2^{\poly(n)}$.

\subsection{Related Work}

\paragraph{Minimal perfect hashing.}
In the early 1980s, Mehlhorn \cite{mehlhorn1982program} and Fredman and Koml\'os \cite{fredman1984size} established that any minimal perfect hash function for $n$ keys requires $n \log e - O(\log n)$ bits, and that this bound can be achieved (if one ignores evaluation time). Schmidt and Siegel \cite{SS90} showed how to achieve $O(1)$-time evaluation with $O(n)$ bits, and Hagerup and Tholey \cite{HT01} subsequently showed how to achieve $O(1)$-time evaluation with $n \log e + o(n)$ bits. There has also been a great deal of work on practical minimal perfect hash functions (see, e.g., \cite{belazzougui2009hash,esposito2020recsplit,lehmann2024shockhash}, the survey by Dietzfelbinger \cite{dietzfelbinger2007design}, and the recent survey by Lehmann et al.\ \cite{lehmann2026modern}), with the most space-efficient constructions now coming within $0.1\%$ of the $n \log e$ lower bound \cite{lehmann2026modern}. A related line of work considers \emph{monotone} minimal perfect hash functions, which must also preserve the order of the keys \cite{belazzougui2009monotone,assadi2023tight,kosolobov2024simplified}.

Many applications of minimal perfect hashing are implicitly instances of value-dynamic retrieval. The value-dynamic retrieval problem was formalized by Kuszmaul and Walzer \cite{kuszmaul2024space}, who proved the $\Omega(n)$ lower bound on redundancy discussed above. Kuszmaul and Walzer also showed how to beat minimal perfect hashing by $\Omega(n / 2^b)$ bits: rather than mapping the keys to an array of size $n$, one can map the keys injectively into an array of size $(1 + \delta)n$ for some appropriately chosen $\delta > 0$. Although the array now has $\delta n$ unused entries, wasting $\delta n b$ bits, the hash function becomes cheaper to encode, and for constant $b$ this is enough to save $\Omega(n)$ bits. Our lower bound implies that, as $b$ grows, the savings from this approach (or from any other approach) must vanish.

\paragraph{Static retrieval.} A closely related problem is that of static retrieval, where queries are allowed but updates are prohibited.
The earliest solutions to the static retrieval problem were implicitly given by minimal perfect hash functions \cite{mehlhorn1982program,fredman1984size,HT01}. The observation that static retrieval can actually be solved with $nb + o(n)$ bits appears to have been made independently by Dietzfelbinger and Pagh \cite{dietzfelbinger2008succinct} and by Porat \cite{porat2009optimal}. Dietzfelbinger and Walzer \cite{dietzfelbinger2019constanttime} later showed how to achieve $O(1)$-time queries with $O(\log n)$ bits of redundancy, in the case where $b = O(1)$. (See also experimental work in \cite{aumuller2009experimental,dillinger2021ribbon,dillinger2022fast}.)

Recently, Hu et al.\ \cite{hu2025static} showed that, when the values are large, static retrieval exhibits an inherent time-space tradeoff: with $t$-time queries in the cell-probe model with $w$-bit words, the optimal redundancy is $n \cdot e^{-\Theta(wt/b)}$ bits (up to lower-order terms, and assuming a polynomial-size universe). For example, if $b = \Theta(\log n)$, then constant-time queries require $\Omega(n)$ bits of redundancy. Note that this time/space tradeoff necessarily also applies to the partially-dynamic setting that we study in the second part of this paper, and indeed, our upper bound in \cref{sec:partially-dynamic-tradeoff} abides by the same time/space tradeoff.

\paragraph{Dynamic retrieval.}
In the other direction, if one wishes to support not just updates to values, but also insertions and deletions of keys, then the optimal redundancy becomes much larger: for such \emph{dynamic} retrieval data structures, with a polynomial-size universe and $b = O(\log n)$, it is known how to achieve $nb + O(n \log \log n)$ bits of space with $O(1)$-time operations \cite{demaine2006dictionariis,kuszmaul2025tight}, even when $n$ is the current number of keys and changes over time \cite{kuszmaul2026resizable}, and there is a matching lower bound of $nb + \Omega(n \log \log n)$ bits for $U \ge n^3$ \cite{kuszmaul2025tight} (building on \cite{chazelle2004bloomier,mortensen2005dynamic}). Interestingly, in the \emph{incremental} setting, where keys can be inserted (and their values updated) but not deleted, the optimal redundancy can be much smaller: for $n^3 \le U \le n^{O(1)}$, it is $\Theta(n + n \log \max\{(\log n)/b, 1\})$ bits \cite{kuszmaul2025tight}, which drops from $\Theta(n \log \log n)$ to $\Theta(n)$ as $b$ approaches $\log n$.

\paragraph{Filters.}
A closely related problem is that of constructing a \emph{filter}, which answers approximate membership queries with a false-positive rate $\epsilon$ (typically assumed to be an inverse power of two). Static filters can be implemented using $n \log \epsilon^{-1} + o(n)$ bits \cite{dietzfelbinger2008succinct,porat2009optimal} (and Hu et al.\ \cite{hu2025static} recently determined the optimal tradeoff between query time and redundancy for static filters), while dynamic filters can be implemented using $n \log \epsilon^{-1} + n \log e + o(n)$ bits \cite{carter1978exact} (with constant-time operations provided that $\log \epsilon^{-1} \in [\omega(1), \log n / \log^{(k)} n]$ for some constant $k$ \cite{bender2022optimal}; see also \cite{pagh2005optimal,arbitman2010backyard,BE20}). Lovett and Porat \cite{lovett2013space} showed that, for $\epsilon = \Theta(1)$, dynamic filters require $n \log \epsilon^{-1} + \Omega(n)$ bits, and Kuszmaul and Walzer \cite{kuszmaul2024space} extended this to all $\epsilon$. Most recently, Kuszmaul, Liang, and Zhou \cite{kuszmaul2025fingerprint} showed that, for $\epsilon = o(1)$, dynamic filters require $n \log \epsilon^{-1} + n \log e - o(n)$ bits, meaning that known constructions are essentially optimal. Other extensions of the problem include filters that support an unknown or growing number of keys \cite{pagh2013how,liu2020succinct}.

Our result can be viewed as the retrieval analogue of \cite{kuszmaul2025fingerprint}: in both settings, the natural hashing-based solution turns out to be optimal, and in both settings, the cost of this solution is $n \log e$ bits beyond the static optimum. The two lower bounds, however, are proven via completely different routes, and appear to come from fundamentally different obstacles. Whether or not there is some unifying approach that somehow capture both bounds at once remains a potentially interesting question.

\section{Preliminaries}
\label{sec:prelim}

\paragraph{Notation.}
Throughout the paper, $\log$ denotes the binary logarithm. For a positive integer $m$, we write $[m] = \{1, 2, \ldots, m\}$. We use $n$ to denote the number of keys, $U$ to denote the size of the universe, and $b$ to denote the number of bits in each value.

\needspace{8\baselineskip}
\paragraph{Problem definitions.}

\begin{definition}[Value-Dynamic Retrieval]
    Let $n,U,b$ be integer parameters where $U\ge n$. In the value-dynamic retrieval problem, the data structure is initially given a set $S\subseteq [U]$ of $n$ keys and a function $f:S\to[2^b]$. After preprocessing the input, the data structure has to support the following operations:
    \begin{itemize}
        \item \textsc{Update}: Given $x\in [U]$ and $v\in [2^b]$, update $f(x)$ to $v$. If $x\notin S$, the behavior of the data structure henceforward is undefined.
        \item \textsc{Query}: Given $x\in [U]$, return $f(x)$. If $x\notin S$, the data structure may answer arbitrarily.
    \end{itemize}
\end{definition}

\begin{definition}[Partially Value-Dynamic Retrieval]
    Let $n,U,b,k$ be integer parameters where $U\ge n$ and $k$ divides $n$. In the partially value-dynamic retrieval problem, the data structure is initially given a set $S\subseteq [U]$ of $n$ keys, a function $f:S\to[2^b]$, and a set $S_{\mathrm{dyn}} \subseteq S$ of $n/k$ \defn{dynamic keys}. We refer to the keys in $S_{\mathrm{stat}} = S \setminus S_{\mathrm{dyn}}$ as \defn{static keys}. The data structure supports \textsc{Query} operations as in value-dynamic retrieval, and supports \textsc{Update} operations only on the dynamic keys: if \textsc{Update} is invoked on any key $x \notin S_{\mathrm{dyn}}$, the behavior of the data structure henceforward is undefined.
\end{definition}

We will sometimes discuss updating the values of a \emph{set} of keys. In this case, we follow the convention that the keys are updated one at a time, in some fixed canonical order (e.g., from smallest to largest).

\paragraph{Computational model.}
Since we are interested in information-theoretic lower bounds, we ignore running time entirely. We refer to the contents of the data structure's memory at a given moment as its \defn{memory configuration}, and we say that the data structure uses $M$ bits of space if every memory configuration that it ever reaches is a string of at most $M$ bits. Critically, the operations must be performed based only on their arguments and on the memory configuration, and not on any additional information about $S$. Thus an \textsc{Update} to some key $x \in S$ must preserve the values of all other keys in $S$, without necessarily having access to $S$ itself.

We assume without loss of generality that the data structure is deterministic, that its memory always consists of exactly $M$ bits, and that \textsc{Update} and \textsc{Query} are defined on \emph{all} inputs, including illegal ones (for completeness, we present in \cref{sec:model-wlog} standard arguments for why these assumptions are without loss of generality).

\paragraph{Encoding arguments.}
All of our lower bounds are proven using one-way communication games, in which Alice holds some random input $X$ and wishes to send it to Bob. Alice and Bob are permitted to share public randomness $R$ that is independent of $X$. We will make use of the following standard fact (see, e.g., \cite{CoverThomas06}).

\begin{fact}
    \label{fact:source-coding}
    If Bob can always recover $X$ from Alice's message and $R$, then the expected length of Alice's message is at least $H(X)$.\footnote{Formally, this requires Alice's messages to form a prefix-free code, so that Bob can tell where a message ends. All of the protocols in this paper can be implemented in this way, with the message lengths that we state.} In particular, if $X$ is uniformly random over a set $\mathcal{X}$, then the expected length is at least $\log |\mathcal{X}|$.
\end{fact}

The lower bounds in this paper are proven in the following way. We pick an input distribution for Alice, whose entropy $H(X)$ we can compute. We then show that a data structure with $M$ bits of space would let Alice send her input with fewer than $H(X)$ bits in expectation, unless $M$ is large. Comparing the two quantities, \cref{fact:source-coding} then gives a lower bound on $M$.

We describe each protocol by listing the parts of Alice's message, together with the number of bits that each part costs. If Bob already knows some set $C$ that contains an element $x$, then Alice can send $x$ using $\lceil \log |C| \rceil$ bits, by sending the rank of $x$ in $C$.

\section{Warm-Up: Lower Bound for Superlinear Value Size}
\label{sec:warm-up}

In this section, we prove a warm-up version of our main result (\cref{thm:warm-up}), for the case where the values are very large. As discussed in \cref{sec:prelim}, we assume without loss of generality that the data structure is deterministic.

\begin{theorem}
    \label{thm:warm-up}
    Let $U = \omega(n)$ and $b = \omega(n)$. Any value-dynamic retrieval data structure that stores $b$-bit values for $n$ keys from $[U]$ uses at least $nb + n \log e - o(n)$ bits of space in the worst case.
\end{theorem}

\subsection{Intuition}
\label{sec:warm-up-intuition}

A lower bound of $nb$ bits is easy to prove: Even if the data structure can store $S$ for free, it still has to remember the $n$ $b$-bit values $\{f(x):x\in S\}$. To improve this to $nb+n\log e-o(n)$ bits, we need to additionally show that the data structure must also store $n\log e-o(n)$ bits of information about $S$.

As an example, it is helpful to think about the data structure that uses minimal perfect hashing. Given a data structure $D$, if we are guaranteed that it works by storing some hash function $h:[U]\to [n]$ and handling operations on $x$ by operating the $h(x)$-th cell, then we can extract a lot of information about $S$ from $D$. Indeed, for any two keys $x,x'\in [U]$, we can detect whether they have the same hash under $h$ by playing with $D$, and if so, they cannot both be in $S$. Doing this for every pair of keys gives us a partition of $[U]$ into $n$ sets $T_1,\dots,T_n$, where two keys are in the same $T_i$ if and only if they have the same hash. Then, $S$ must contain exactly one key from each set, and thus the number of possible $S$'s given $D$ is at most
\begin{align*}
    \prod_{i=1}^{n}|T_i|\le \bk*{\frac Un}^n.\tag{Jensen's inequality}
\end{align*}
In comparison, the total number of size-$n$ subsets of $[U]$ is $\binom{U}{n}$. Thus, conditioning on $D$ reduces the entropy of $S$ by at least
\begin{align*}
    &\log\binom{U}{n}-\log\bk*{\bk*{\frac Un}^n} \\
    ={}&n\log\bk*{\frac{U}{n}}+n\log e-o(n)-n\log\bk*{\frac{U}{n}}\tag{Stirling's formula, as $U = \omega(n)$} \\
    ={}&n\log e-o(n).
\end{align*}
Since the $n$ values are independent of $S$, and $D$ must also determine all of them, $D$ must therefore contain at least $nb + n \log e - o(n)$ bits of information, which is the bound that we want.

The idea for the general lower bound is similar but slightly more complicated. We also consider every pair of keys $x,x'\in [U]$, and see if updating $x$ would change the value of $x'$. Similarly to before, any pair of keys in the real set $S$ must not corrupt each other. Unlike the previous case, this relation may not just be a partition of $[U]$. However, we can still bound the number of candidates for $S$. There are two cases:

\begin{itemize}
    \item If there exists some $x\notin S$ that does not corrupt any key in $S$, then intuitively, our data structure is secretly a retrieval data structure for $n+1$ keys. This means that it must use $(n+1)b$ bits of space, which when $b=\omega(n)$, is already sufficient for us.
    \item Otherwise, any $x\notin S$ corrupts some key in $S$. In this case, writing $S = \{x_1, \ldots, x_n\}$, we similarly consider a partitioning of keys into $n$ sets: A key $x\in [U]$ is placed in $T_i$ if $x$ corrupts $x_i$ (if there are multiple such $x_i$, then we place $x$ arbitrarily in one of them). Similarly to before, the real set $S$ contains one key from each $T_i$. However, this time Bob cannot compute $T_i$ on his own, because this would require knowing $S$.

    We get around this by considering a communication game: Suppose Alice wants to send $S=\{x_1,\dots,x_n\}$ to Bob (here we think of $S$ as an ordered set), and Bob already knows $D$: Alice randomly permutes the keys in $S$ into $x_{\pi_1},\dots,x_{\pi_n}$, and sends them to Bob sequentially, where each $x_{\pi_i}$ is encoded optimally given $x_{\pi_1},\dots,x_{\pi_{i-1}}$. After seeing $x_{\pi_1},\dots,x_{\pi_{i-1}}$, Bob can compute the set $C_i$ of keys that do not corrupt them. By definition of $T_i$, $C_i$ cannot contain any key in $T_{\pi_j},j<i$. Therefore,
    \begin{align*}
        |C_i|\le U-\sum_{j=1}^{i-1}|T_{\pi_j}|.
    \end{align*}
    Since we must have $x_{\pi_i}\in C_i$, Alice's message has expected length
    \begin{align*}
        &\E\Bk*{\sum_{i=1}^{n}\log |C_i|}\le\E\Bk*{\sum_{i=1}^{n}\log \bk*{U-\sum_{j=1}^{i-1}|T_{\pi_j}|}} \\
        \le{}&\sum_{i=1}^{n}\log\E\Bk*{U-\sum_{j=1}^{i-1}|T_{\pi_j}|}\tag{Jensen's inequality} \\
        \le{}&\sum_{i=1}^{n}\log \bk*{U\cdot\frac{n-i+1}{n}}=\log\bk*{\bk*{\frac{U}{n}}^n}+\log (n!).
    \end{align*}
    (For the last inequality, note that the sets $T_1, \ldots, T_n$ partition $[U]$, and that a random permutation picks $i - 1$ of them, so the expected total size of the picked sets is $\frac{i - 1}{n} \cdot U$.) In comparison, $S$ (as an ordered set) has entropy $\log(U!/(U-n)!)=\log\binom{U}{n}+\log (n!)$. Therefore, we save $n\log e-o(n)$ bits as before.
\end{itemize}

This intuition glosses over two details that the formal proof must handle. First, in the first case, $x$ must do more than leave the keys in $S$ uncorrupted. It must also behave like a real key itself, so that the extra value can be read back out after it is written. This is why the notion of consistency that we define in \cref{sec:tools} includes a requirement that we call internal consistency. Second, in the second case, we are aiming to save only $n \log e$ bits, so losing even one bit of rounding per key would be too costly. This is why Alice encodes the keys jointly, using \cref{lem:sequential} below, rather than one at a time.

\subsection{Notation and Tools}
\label{sec:tools}

\paragraph{Consistency.}
The following definitions capture which real keys get corrupted when we update the values of keys that are \emph{not} in $S$. Fix a memory configuration $D$ of a (partially) value-dynamic retrieval data structure, and, for a set $A \subseteq [U]$, define an \defn{update sequence on $A$} to be a finite sequence of \textsc{Update} operations, each of which is on some key in $A$. Recall from \cref{sec:prelim} that we assume that \textsc{Update} and \textsc{Query} are defined on all inputs. So it makes sense to apply any update sequence to $D$, even if $A \not\subseteq S$.

\begin{definition}
    \label{def:consistency}
    Let $A \subseteq [U]$ be a set of keys.
    \begin{itemize}
        \item $A$ is \defn{internally consistent} (on $D$) if, for every update sequence on $A$, applying the sequence to $D$ results in a memory configuration in which every $y \in A$ is answered (by \textsc{Query}) with the value most recently assigned to $y$ by the sequence, or, if the sequence never updates $y$, with the same answer as on $D$.
        \item For a key $x \in [U]$, $A$ is \defn{consistent w.r.t.\ $x$} (on $D$) if $A$ is internally consistent, and if every update sequence on $A$, applied to $D$, leaves the answer to $\textsc{Query}(x)$ unchanged. Otherwise, $A$ is \defn{inconsistent w.r.t.\ $x$}.
        \item $A$ is consistent w.r.t.\ a set of keys if it is consistent w.r.t.\ every key in that set. A single key $y$ is consistent w.r.t.\ $x$ if $\{y\}$ is.
    \end{itemize}
\end{definition}

In other words, $A$ is consistent w.r.t.\ $x$ if the data structure can store values in $A$ without disturbing $x$. We point out two simple consequences of the definition, which we will use repeatedly. First, a set is never consistent w.r.t.\ one of its own keys: if $x \in A$, then updating $x$ to a value that differs from its current answer changes the answer to $\textsc{Query}(x)$. Second, internal consistency is exactly what allows us to use $A$ as extra storage: if $A$ is internally consistent, then any values written to the keys of $A$ can later be read back by querying them. We will in several places make use of the following simple observation.

\begin{observation}
    \label{obs:legal-consistent}
    Let $D$ be the memory configuration of a value-dynamic retrieval data structure that has been constructed on the key set $S$. Then every set $A \subseteq S$ is consistent w.r.t.\ every key $x \in S \setminus A$. Similarly, for a partially value-dynamic retrieval data structure constructed with dynamic keys $S_{\mathrm{dyn}}$, every set $A \subseteq S_{\mathrm{dyn}}$ is consistent w.r.t.\ every key $x \in S \setminus A$.
\end{observation}

Indeed, every update sequence on such a set $A$ is a legal sequence of operations, so, after applying it to $D$, the data structure must answer every query to a key in $S$ correctly. In particular, each $y \in A$ is answered with the value most recently assigned to it (or with its original value, if it was never updated), and $x$ is answered with its original value.

\paragraph{Encoding sequences without rounding losses.}
To send a sequence of elements, each of which comes from a set that depends on the previous elements, encoding the elements separately would lose up to one bit of rounding per element, which is too much for our purposes. (For example, in \cref{sec:warm-up-proof}, Alice will send $n$ keys and will aim to save about $n \log e \approx 1.44 n$ bits in total, so she cannot afford to lose up to one bit per key.) Instead, Alice can encode the entire sequence as a single integer written in mixed radix:

\begin{lemma}
    \label{lem:sequential}
    Suppose that Alice wishes to send a sequence $z_1, \ldots, z_\ell$ to Bob, and that for each $i \in [\ell]$ there is a set $C_i \ni z_i$ that Bob can compute from $z_1, \ldots, z_{i - 1}$ (along with information that he already knows, including $\ell$). Then Alice can send $z_1, \ldots, z_\ell$ using
    $$\sum_{i = 1}^\ell \log |C_i| + O\left(\log \left(2 + \sum_{i = 1}^\ell \log |C_i|\right)\right)$$
    bits.
\end{lemma}
\begin{proof}
    Letting $r_i$ be the rank of $z_i$ in $C_i$ (starting from $0$), Alice sends $L = \lceil \sum_i \log |C_i| \rceil$, using $O(\log(L + 2))$ bits, followed by the integer $N = \sum_{i} r_i \prod_{j < i} |C_j| < \prod_i |C_i|$ written using $L$ bits. For $i = 1, 2, \ldots, \ell$, Bob computes $C_i$ and recovers $r_i = \lfloor N / \prod_{j < i} |C_j| \rfloor \bmod |C_i|$, and hence $z_i$.
\end{proof}

Intuitively, Alice writes the ranks $r_1, \ldots, r_\ell$ as the digits of a single number $N$, where the $i$-th digit (counting from the least significant one) is written in base $|C_i|$. Bob reads off the digits starting from the least significant one. By the time that he needs the base $|C_i|$ of the $i$-th digit, he has already decoded $z_1, \ldots, z_{i - 1}$, so he can compute $C_i$. Since $N < \prod_i |C_i|$, writing $N$ costs only $\lceil \sum_i \log |C_i| \rceil$ bits, so the total loss due to rounding is at most one bit, rather than one bit per element.

In our applications of \cref{lem:sequential}, we will always have $\sum_i \log |C_i| \le n \log U = O(n \log n)$, so the additive overhead will be $O(\log n)$ (here we assume that $U \le n^2$, which is without loss of generality: a data structure for the universe $[U]$ is also one for the universe $[\min(U, n^2)]$, and each of our lower bounds only requires $U = \omega(n)$ or $U \ge 2n$).

\paragraph{Counting sequences of distinct keys.}
For integers $U \ge n \ge 0$, note that $U!/(U - n)!$ is the number of sequences of $n$ distinct keys from $[U]$. By Stirling's approximation, we have $\log (n!) = n \log n - n \log e + O(\log n)$. We will also use the fact that, if $U = \omega(n)$, then
\begin{equation}
    \log \frac{U!}{(U-n)!} = n \log U - o(n),
    \label{eq:ff}
\end{equation}
which follows from $n \log (U - n) \le \log \frac{U!}{(U-n)!} \le n \log U$ and $n \log(U - n) = n \log U + n \log(1 - n/U) = n \log U - O(n^2/U)$. In words, when $U \gg n$, requiring the $n$ keys to be distinct barely reduces the number of possibilities below $U^n$. So we can pretend that the keys are chosen independently, while losing only $o(n)$ bits.

\subsection{The Lower Bound}
\label{sec:warm-up-proof}

\paragraph{The communication game.} We formally describe the proof using a communication protocol. Consider the one-way communication game where Alice knows $n$ \emph{distinct} keys $x_1,\dots,x_n\in [U]$ as well as $n+1$ values $v_1,\dots,v_{n+1}\in [2^b]$, all chosen uniformly at random, and wishes to send them to Bob. The information-theoretic lower bound of Alice's message is
\begin{align*}
    \log \frac{U!}{(U-n)!}+(n+1)b
\end{align*}
bits (note that while the keys $x_1,\dots,x_n$ are distinct, they are ordered). By \cref{fact:source-coding}, this is a lower bound on the expected length of Alice's message in any protocol, since her input is uniformly random over a set of $\frac{U!}{(U-n)!} \cdot 2^{(n+1)b}$ possibilities.

In the following, we show that if there exists a value-dynamic retrieval data structure using $M$ bits, then Alice can send the information to Bob using at most
\begin{align}
    \label{equ:warm-up-goal}
    M+\log\frac{U!}{(U-n)!}+b-n\log e+o(n)
\end{align}
bits in expectation. As a result, by \cref{fact:source-coding}, we must have $M\ge nb+n\log e-o(n)$. (Indeed, \eqref{equ:warm-up-goal} must be at least $\log \frac{U!}{(U-n)!}+(n+1)b$. The terms $\log \frac{U!}{(U-n)!}$ and $b$ cancel, which leaves $M - n \log e + o(n) \ge nb$.)

\paragraph{Definitions.} Let $S = \{x_1, \ldots, x_n\}$ be the real key set, and let $D$ denote the memory configuration of the data structure initialized with $(x_1,v_1),$ $\dots$ $,(x_n,v_n)$. We use the notion of consistency from \cref{def:consistency}. Recall that, for keys $x,x'\in [U]$, $x'$ is consistent w.r.t.\ $x$ (on $D$) if, starting from $D$, we can arbitrarily update the value of $x'$, and the answer to $\textsc{Query}(x')$ will always be the most recently assigned value, and the answer to $\textsc{Query}(x)$ will not change. In particular, $x$ is not consistent w.r.t.\ itself.

We say that a key $x\in [U]$ is \defn{deceptive w.r.t. $\{x_i\}_{1\le i\le n}$ (on $D$)}, if $x$ is consistent w.r.t. every $x_i$ for $1\le i\le n$. Note that a deceptive key cannot be one of $x_1, \ldots, x_n$. In other words, a deceptive key can be updated without corrupting any real key, and (by internal consistency) the value written to it can be read back. This is the formal version of a key ``$x \notin S$ that does not corrupt any key in $S$'' from the intuition.

Our protocol considers two cases: Either there exists a deceptive key, or there does not.

\paragraph{Case 1: There exists a deceptive key.} Intuitively, this is the case where our data structure is secretly a retrieval data structure for $n+1$ keys. Let $y$ denote the smallest deceptive key. Let $D'$ denote the data structure obtained from $D$ by executing the operation $\textsc{Update}(y,v_{n+1})$. Alice sends the following message to Bob:

\begin{itemize}
    \item One bit indicating that we are in case 1.
    \item The data structure $D'$, using $M$ bits.
    \item The keys $x_1,\dots,x_n$, using $\lceil \log\frac{U!}{(U-n)!} \rceil$ bits.
\end{itemize}

\paragraph{Correctness.} After receiving the message, Bob can recover $v_1,\dots,v_n$ by querying $D'$. Since $y$ is consistent w.r.t. every $x_i$ on $D$, the answers to $\textsc{Query}(x_i)$ are unchanged when we evolve from $D$ to $D'$.

After this, Bob can simulate the construction of $D$, and recover $y$ from $D$ and $x_1,\dots,x_n$ by its definition. Then, Bob can query $y$ on $D'$ to recover $v_{n+1}$ (here we use that $y$ is internally consistent, so $\textsc{Query}(y)$ returns the value $v_{n+1}$ that was most recently assigned to it). Therefore, this protocol is correct.

\paragraph{Length of message.} The total length of the message is at most
\begin{align*}
    &2+M+\log\frac{{U!}}{(U-n)!}
\end{align*}
bits, which is always smaller than \eqref{equ:warm-up-goal} (here we use the fact that $b\gg n\log e$, which follows from $b = \omega(n)$). Intuitively, the extra value $v_{n+1}$ is sent for free, since it is hidden inside $D'$.

\paragraph{Case 2: There is no deceptive key.} In this case, Alice sends the keys $x_1,\dots,x_n$ in a random order, where each key is encoded optimally based on previous keys. Formally, Alice sends the following message to Bob:

\begin{itemize}
    \item One bit indicating that we are in case 2.
    \item The data structure $D$, using $M$ bits.
    \item The extra value $v_{n+1}$, using $b$ bits.
    \item Sample using public randomness a permutation $\pi:[n]\to[n]$ (so Bob knows $\pi$ for free), and send the keys $x_{\pi_1},\dots,x_{\pi_n}$ using \cref{lem:sequential}, where the set $C_i$ for $x_{\pi_i}$ is defined below.
\end{itemize}

\paragraph{Correctness.} After receiving $x_{\pi_1},\dots,x_{\pi_n}$, Bob can permute the keys back to the original order, and query the keys on $D$ to recover the values $v_1,\dots,v_n$.

\paragraph{Length of message.} In the last step, for each $1\le i\le n$, Alice encodes $x_{\pi_i}$ optimally conditioned on $D$ and $x_{\pi_1},\dots,x_{\pi_{i-1}}$. In particular, after receiving the previous message, Bob can compute $C_i$, the set of keys that are consistent w.r.t. $x_{\pi_1},\dots,x_{\pi_{i-1}}$ on $D$. We have $x_{\pi_i}\in C_i$ by \cref{obs:legal-consistent} (which says that updating real keys never corrupts the other real keys), applied to the set $\{x_{\pi_i}\}$. So, by \cref{lem:sequential} and the discussion after it, Alice's encoding of all $n$ keys costs at most $\sum_i \log|C_i| + O(\log n)$ bits. Since there is no deceptive key, every $x\in [U]$ is inconsistent w.r.t. some key in $\{x_1,\dots,x_n\}$ (call it $x^*$). As a result, the probability of $x$ contributing to $C_i$ is at most $(n-i+1)/n$ (the probability that $x^*$ did not appear in the first $i-1$ keys under $\pi$). Thus, by Jensen's inequality,

\begin{align*}
    \E[\log |C_i|]\le \log\E[|C_i|]\le \log \bk*{U\cdot \frac{(n-i+1)}{n}},
\end{align*}
and the total expected length of Alice's message is
\begin{align*}
    &\le 1+M+b+\sum_{i=1}^{n}\log \bk*{U\cdot\frac{(n-i+1)}{n}} + O(\log n) \\
    ={}&M+b+\log\bk*{U^n\cdot\frac{n!}{n^n}} + O(\log n) \\
    ={}&M+b+n \log U - n\log e+O(\log n)\tag{Stirling's formula} \\
    \le{}&M+\log\frac{U!}{(U-n)!}+b-n\log e+o(n)\tag{by \eqref{eq:ff}, since $U = \omega(n)$},
\end{align*}
which fits \eqref{equ:warm-up-goal}. This concludes the proof of \cref{thm:warm-up}.

\section{Lower Bound for Superconstant Value Size}
\label{sec:super-constant-v}

In this section, we generalize the warm-up to handle arbitrary superconstant value sizes.

\begin{theorem}
    \label{thm:super-constant-v}
    Let $U = \omega(n)$. Consider a value-dynamic retrieval data structure that stores $b$-bit values for $n$ keys from $[U]$. If $b=\omega(1)$, then the data structure uses at least $nb+(\log e)\cdot n-o(n)$ bits of space in the worst case.
\end{theorem}

\paragraph{Intuition.} In the warm-up, the protocol branched on whether there exists a deceptive key (a key that can be updated without corrupting any real key, and whose value can be read back). When $b$ is small, a single deceptive key is no longer worth $n \log e$ bits, since it only hides one extra $b$-bit value. So instead we look for \emph{sets} of $s \approx n/g$ keys, for some parameter $g$, that can collectively store $sb$ bits of extra information. The most subtle part of the proof is finding the right definition of a deceptive \emph{set}, so that the protocol saves $n \log e - o(n)$ bits regardless of whether such a set exists.

The most natural definition would be to call a set of $s$ keys deceptive if it is internally consistent and does not corrupt any real key in $S$. If such a set exists, then, just as in the warm-up, Alice can store $sb$ bits of extra information in it, which is more than $n \log e$ as long as $g$ is small compared to $b$ (since $sb \approx nb/g$). If no such set exists, however, then the natural analogue of the warm-up protocol is for Alice to send the keys in $g$ random \emph{batches} of $s$ keys each, where Bob uses the keys in the first $j - 1$ batches to rule out candidates for the $j$-th batch. But a non-deceptive set might corrupt only a single real key, in which case it survives as a candidate for the $j$-th batch with probability $1 - (j - 1)/g$ (the probability that this one real key is not among the roughly $\frac{j - 1}{g} \cdot n$ keys in the first $j - 1$ batches). This lets Alice save only
$$\sum_{j = 1}^{g} \log \frac{1}{1 - (j - 1)/g} = \log \frac{g^g}{g!} = g \log e - O(\log g)$$
bits in total, far less than the $n \log e$ bits that we need. Here, the $j$-th term is the savings on the $j$-th batch: just as in the warm-up, if each candidate survives with probability $1 - (j - 1)/g$, then the expected number of candidates shrinks by this factor, and Alice saves $\log \frac{1}{1 - (j - 1)/g}$ bits. The terms add up to $\log \prod_{j = 1}^{g} \frac{g}{g - j + 1} = \log \frac{g^g}{g!}$, which is about $g \log e$ by Stirling's approximation. The warm-up is the special case $s = 1$ and $g = n$, where this gives the desired $n \log e$ bits. With batches of $s$ keys, we lose a factor of $s$.

Alternatively, we could call a set deceptive if it corrupts at most $s$ real keys. If no such set exists, then every set corrupts more than $s$ real keys, and therefore survives as a candidate for the $j$-th batch with probability only about $(1 - (j - 1)/g)^s$ (since each of the more than $s$ real keys that it corrupts must avoid the first $j - 1$ batches). This multiplies the savings by a factor of $s$, giving $s \cdot (g \log e - o(g)) = n \log e - o(n)$ bits, as desired. However, with this definition, the case where a deceptive set \emph{does} exist no longer works, since the deceptive set may corrupt as many as $s$ real keys, and the $sb$ bits needed to send their values explicitly cancel out the $sb$ bits that we gained.

To get the best of both worlds, we introduce a parameter $\varepsilon$, and call a set deceptive if it is internally consistent and corrupts at most $(1 - \varepsilon) s$ real keys. If a deceptive set exists, then Alice gains $sb$ bits and loses at most $(1 - \varepsilon) sb$ bits (for sending the values of the corrupted real keys explicitly), for a net savings of roughly $\varepsilon s b$ bits. If no deceptive set exists, then the argument above saves roughly $(1 - \varepsilon) s \cdot g \log e \approx (1 - \varepsilon) n \log e$ bits. For example, if we set $g \approx \sqrt{b}$ and $\varepsilon = b^{-1/3}$, then the first case saves roughly $\varepsilon s b \approx b^{-1/3} \cdot \frac{n}{\sqrt{b}} \cdot b = b^{1/6} n$ bits, and the second case saves roughly $(1 - b^{-1/3}) n \log e$ bits. Both are at least $n \log e - o(n)$ when $b = \omega(1)$.

This high-level intuition ignores a few details that the formal proof must handle. In the first case, Alice must also tell Bob \emph{which} real keys were corrupted, which costs roughly $s \log g$ bits. This is why we need $\varepsilon b$ to be large compared to $\log g$. In the second case, $n$ need not be a multiple of $g$, so Alice sends the leftover keys explicitly. Additionally, to avoid losing a bit of rounding on each batch, Alice sends all of the batches jointly using \cref{lem:sequential} (the mixed-radix encoding from \cref{sec:tools}).

\paragraph{A more general statement.}
It will be convenient to prove a more general statement, which is parameterized by $g$ and $\varepsilon$. We will use this statement both to prove \cref{thm:super-constant-v} and, in \cref{sec:constant-v}, to obtain a lower bound whose dependence on $b$ is explicit.

\begin{proposition}
    \label{prop:general}
    Let $U = \omega(n)$, let $g$ be an integer with $2 \le g \le \sqrt{n}$, let $s = \lfloor n/g \rfloor$, and let $\varepsilon \in (0, 1/2]$. Consider a value-dynamic retrieval data structure that stores $b$-bit values for $n$ keys from $[U]$ using $M$ bits of space. Then
    $$M \ge nb + \min\left\{\varepsilon s b,\ (1 - \varepsilon) n \log e\right\} - O\left(\frac{n \log g}{g}\right) - o(n).$$
\end{proposition}

The two terms in the minimum are the savings in the two cases of the protocol sketched above. In the first case, each of the $s$ keys of a deceptive set saves $\varepsilon b$ bits. In the second case, Alice saves by sending the keys in batches. The error term accounts for describing which real keys a deceptive set corrupts (about $\log g$ bits per key of the set), and for the lower-order terms in Stirling's approximation. To obtain a strong bound, we want $\varepsilon$ to be small and $g$ to be large, but not so large that $\varepsilon s b \approx \varepsilon n b / g$ drops below $n \log e$.

We first show how \cref{prop:general} implies \cref{thm:super-constant-v}.

\begin{proof}[Proof of \cref{thm:super-constant-v}]
    Set $g = \lfloor \min(b, n)^{1/2} \rfloor$ and $\varepsilon = b^{-1/3}$, and assume that $n$ is sufficiently large. Since $b = \omega(1)$, we have $g = \omega(1)$ and $\varepsilon = o(1)$; in particular, $2 \le g \le \sqrt{n}$ and $\varepsilon \le 1/2$. Since $s \ge n/g - 1 \ge n/(2g)$ and $g \le \sqrt{b}$, the first term in the minimum is
    $$\varepsilon s b \ge \frac{n}{2} \cdot \frac{\varepsilon b}{g} \ge \frac{n}{2} \cdot b^{1/6} \ge n \log e.$$
    The second term is $(1 - \varepsilon) n \log e = n \log e - o(n)$, and the error term $O(\frac{n \log g}{g})$ is $o(n)$ since $g = \omega(1)$. So \cref{prop:general} gives $M \ge nb + n \log e - o(n)$.
\end{proof}

The rest of the section proves \cref{prop:general}.

\paragraph{The communication game.}
Let $g$, $s$, and $\varepsilon$ be as in \cref{prop:general}. Consider the one-way communication game where Alice sends to Bob an ordered sequence of $n$ \emph{distinct} keys $x_1,\dots,x_n\in [U]$, as well as $n+s$ values $v_1,\dots,v_{n+s}\in [2^b]$, all chosen uniformly at random. The information-theoretic optimal message length is
\begin{align*}
    \log \frac{U!}{(U-n)!} + (n + s)\cdot b.
\end{align*}

In the following, we show that, if there exists a value-dynamic retrieval data structure that always uses at most $M$ bits of space, then we can send the information using
\begin{align}
    \label{equ:goal}
    \log \frac{U!}{(U-n)!} + s\cdot b + M - \min\{\varepsilon s b,\ (1 - \varepsilon) n \log e\} + O\left(\frac{n \log g}{g}\right) + o(n)
\end{align}
bits of communication in expectation. By \cref{fact:source-coding}, this proves \cref{prop:general} (the expected message length must be at least the entropy $\log \frac{U!}{(U-n)!} + (n + s) b$ of Alice's input, and, after cancelling the common terms, this is exactly the claimed bound on $M$).

\paragraph{Definitions.} Let $S=\{x_1,\dots,x_n\}$ be the real key set, and let $D$ denote the memory configuration of the data structure initialized with $(x_1,v_1),$ $\dots$ $,(x_n,v_n)$. We use the notions of consistency from \cref{def:consistency}.

Given a set $T$ of $s$ keys, we say that it is \defn{deceptive w.r.t. $S$ (on $D$)}, if for all but at most $(1-\varepsilon)\cdot s$ keys $x\in S$, $T$ is consistent w.r.t. $x$. Intuitively, a deceptive set can store $s$ extra values while corrupting at most $(1 - \varepsilon) s$ real keys. Note that a deceptive set is automatically internally consistent. Otherwise, since consistency w.r.t.\ a key requires internal consistency (\cref{def:consistency}), $T$ would be inconsistent w.r.t.\ all $n > (1 - \varepsilon) s$ real keys.

Our protocol considers two cases: Either there exists a deceptive set, or there does not.

\paragraph{Case 1: There exists a deceptive set.} Let $Y=\{y_1,\dots,y_{s}\}$ denote the first deceptive set (under some canonical order of sets), where $y_1 < \cdots < y_s$. Let $D'$ denote the data structure obtained from $D$ by executing the operations $\textsc{Update}(y_i,v_{n+i})$ for every $1\le i\le s$. Let $X_{\text{bad}}$ denote the set of indices $i$ such that $Y$ is not consistent w.r.t.\ $x_i$ (on $D$), and note that $|X_{\text{bad}}| \le (1 - \varepsilon) s$. Alice sends the following message to Bob:

\begin{itemize}
    \item One bit indicating that we are in case 1.
    \item The data structure $D'$, using $M$ bits.
    \item The keys $x_1,\dots,x_n$, using $\lceil \log \frac{U!}{(U-n)!} \rceil$ bits.
    \item The set $X_{\text{bad}}$, encoded as one of the at most $(en/s)^s$ subsets of $[n]$ of size at most $s$, using $\lceil s \log(en/s) \rceil$ bits.
    \item The input values associated with the keys in $X_{\text{bad}}$ (i.e., if $i\in X_{\text{bad}}$, we send $v_i$). This uses $|X_{\text{bad}}| \cdot b \le (1 - \varepsilon) s \cdot b$ bits.
\end{itemize}

\paragraph{Correctness.} After receiving the message, Bob can recover $v_1,\dots,v_n$: For a key $x_i$ with $i \notin X_{\text{bad}}$, the updates on $Y$ do not change its value, so Bob can recover $v_i$ by querying on $D'$. For the keys with indices in $X_{\text{bad}}$, their values are explicitly given.

Bob can also simulate the construction of $D$, and recover $Y$ from $D$ and $x_1,\dots,x_n$. Then, Bob can query the keys in $Y$ on $D'$ to recover $v_{n+1},\dots,v_{n+s}$ (here we use that $Y$ is internally consistent). Therefore, this protocol is correct.

\paragraph{Length of message.} Since $n/s \le 2g$, the fourth step costs at most $s \log(2eg) + 1 = O(\frac{n \log g}{g})$ bits: each of the at most $s$ corrupted keys is one of $n \approx gs$ real keys, so describing it costs about $\log g$ bits. Indeed, the fourth and fifth steps are the cost of erasing some information in the real set $S$, which as we discussed in the intuition is roughly $(1-\varepsilon)\cdot s\cdot b$ bits. In total, the message length is at most
\begin{align*}
    M + \log \frac{U!}{(U-n)!} + (1 - \varepsilon) s b + O\left(\frac{n \log g}{g}\right),
\end{align*}
which is at most \eqref{equ:goal}, since the minimum in \eqref{equ:goal} is at most $\varepsilon s b$.

\paragraph{Case 2: There is no deceptive set.} In this case, Alice sends the keys in batches of $s$, where each batch is encoded optimally based on previous batches. The intuition is that, once Bob has seen a key $x$, he knows that Alice will not send any key set that is inconsistent w.r.t.\ $x$. This helps reduce the cost of communication. Since $n$ need not be a multiple of $g$, Alice first sends the $r = n - gs < g$ keys that do not fit into any batch explicitly. Formally, Alice sends the following message to Bob:

\begin{itemize}
    \item One bit indicating that we are in case 2.
    \item The data structure $D$, using $M$ bits.
    \item The extra values $v_{n+1},\dots,v_{n+s}$, using $s\cdot b$ bits.
    \item Sample using public randomness a permutation $\pi:[n]\to[n]$. Alice sends the keys $x_{\pi_{gs + 1}}, \ldots, x_{\pi_n}$, followed by the batches $X_{\pi, 1}, \ldots, X_{\pi, g}$, where, for $1\le j\le g$, $X_{\pi,j}$ is the (\emph{ordered}) set
    \begin{align*}
        X_{\pi,j}\coloneqq x_{\pi_{s\cdot (j-1)+1}},\dots,x_{\pi_{s\cdot j}}.
    \end{align*}
    All of these are sent together using \cref{lem:sequential}. Each of the first $r$ keys is encoded as an element of $[U]$, and each batch $X_{\pi, j}$ is encoded as an element of the set $C_j$ defined below.
\end{itemize}

\paragraph{Correctness.} After receiving $x_{\pi_1},\dots,x_{\pi_n}$, Bob can permute the keys back to the original order, and query the keys on $D$ to recover the values $v_1,\dots,v_n$.

\paragraph{Analysis for the last step.} For $j \in [g]$, let $P_j \subseteq [n]$ denote the set of indices of the keys that Bob knows prior to receiving $X_{\pi, j}$, that is, $P_j = \{\pi_i \mid i > gs \text{ or } i \le s(j - 1)\}$. Let $C_j$ denote the set of ordered sequences of $s$ distinct keys in $[U]$ that are consistent w.r.t.\ every key $x_i$ with $i \in P_j$ (where we say that a sequence is consistent w.r.t.\ a key if the set of its keys is). Bob can compute $C_j$ from $D$ and the keys that he has already received. Moreover, by \cref{obs:legal-consistent}, which says that updating real keys never corrupts the other real keys, the keys in $X_{\pi, j}$ are consistent w.r.t.\ all of the other keys in $S$, so $X_{\pi, j} \in C_j$. It remains to show that $|{C_{j}}|$ is small. By Jensen's inequality,
\begin{align*}
    \E[\log |C_{j}|]&\le \log \E[|C_{j}|] \\
    &=\log\bk*{\sum_{\text{ordered } T \subseteq [U],\, |T| = s}\Pr[T\text{ is consistent w.r.t. all keys $x_i$ with $i \in P_j$}]}.
\end{align*}
Fix such a $T$, and let $I \subseteq [n]$ denote the set of indices $i$ such that $T$ is inconsistent w.r.t.\ $x_i$. Since no set is deceptive, we have $|I| > (1 - \varepsilon) s$, and $T$ is consistent w.r.t.\ all keys $x_i$ with $i \in P_j$ if and only if $I \cap P_j = \emptyset$. The set $P_j$ is a uniformly random subset of $[n]$ of size $r + (j - 1)s$, and
$$\frac{r + (j - 1)s}{n} = \frac{r + (j - 1)s}{gs + r} \ge \frac{j - 1}{g}.$$
In words, Bob already knows at least a $\frac{j-1}{g}$ fraction of the real keys, so $T$ survives only if all of the more than $(1 - \varepsilon) s$ real keys that it corrupts are among the keys that Bob does not know yet. To bound the probability of this, we use the following simple fact.

\begin{fact}
    \label{fact:random-subset}
    Let $R$ be a uniformly random subset of $[n]$ of some fixed size $m$, and let $I \subseteq [n]$ be a fixed set of size $\ell$. Then $\Pr[I \cap R = \emptyset] \le (1 - m/n)^{\ell}$.
\end{fact}

Indeed, the probability is $\prod_{i = 0}^{\ell - 1} \frac{n - m - i}{n - i}$, and each factor is at most $(n - m)/n$. In other words, the elements of $I$ behave no worse than if each of them were included in $R$ independently with probability $m/n$.

Thus, by \cref{fact:random-subset}, applied with $R = P_j$,
\begin{align*}
    \Pr[I \cap P_j = \emptyset] \le \left(1 - \frac{j - 1}{g}\right)^{|I|} \le \bk*{\frac{g-j+1}{g}}^{(1-\varepsilon)\cdot s}.
\end{align*}
This means that
\begin{align*}
    \E[\log |C_{j}|]\le \log\bk*{\frac{U!}{(U-s)!}\cdot \bk*{\frac{g-j+1}{g}}^{(1-\varepsilon)\cdot s}}.
\end{align*}
Here, $\frac{U!}{(U-s)!}$ is the number of ordered sets $T$ with $T\subseteq [U],|T|=s$.

In total, by \cref{lem:sequential}, the expected cost of the last step is at most
\begin{align*}
    &r \log U + \sum_{j=1}^{g}\E[\log |C_{j}|] + O(\log n) \\
    \le{}&r \log U + \sum_{j=1}^{g}\bk*{\log \frac{U!}{(U-s)!} + (1-\varepsilon)\cdot s\cdot \log\bk*{\frac{g-j+1}{g}}} + O(\log n) \\
    \le{}&(r + gs) \log U - (1-\varepsilon)\cdot s\cdot \log\bk*{\frac{g^g}{g!}} + O(\log n),
\end{align*}
where the second inequality uses $\frac{U!}{(U-s)!} \le U^s$ and $\sum_{j = 1}^{g} \log \frac{g - j + 1}{g} = \log \frac{g!}{g^g}$. By Stirling's approximation, $\log \frac{g^g}{g!} = g \log e - O(\log g)$. Since $r + gs = n$ and $gs \ge n - g$, and since $g \le \sqrt{n}$, the expected cost of the last step is therefore at most
$$n \log U - (1 - \varepsilon) n \log e + O\left(\frac{n \log g}{g}\right) + o(n).$$

\paragraph{Length of message.} The total expected length of the message is at most
\begin{align*}
    1 + M + s b + n \log U - (1 - \varepsilon) n \log e + O\left(\frac{n \log g}{g}\right) + o(n),
\end{align*}
which, by \eqref{eq:ff}, is at most \eqref{equ:goal}, since the minimum in \eqref{equ:goal} is at most $(1 - \varepsilon) n \log e$.

In conclusion, the expected message length of our protocol is always at most \eqref{equ:goal}. Comparing \eqref{equ:goal} with the information-theoretic lower bound of $\log \frac{U!}{(U-n)!} + (n + s) b$ bits proves \cref{prop:general}.

\section{Lower Bound for Small Value Sizes}
\label{sec:constant-v}

In this section, we give a lower bound on the redundancy of value-dynamic retrieval data structures whose dependence on $b$ is explicit (\cref{thm:constant-v}): the redundancy is at least $(\log e - O(\sqrt{(\log b)/b}))\, n - o(n)$ bits. This shows that the optimal redundancy approaches $n \log e$ as $b$ grows, and gives a quantitative bound on how fast. In particular, this bound is meaningful even when $b$ is a sufficiently large constant, where \cref{thm:super-constant-v} does not apply. \Cref{prop:general} still applies for any $b$, but the parameter $g$ is now limited by $b$, so we can no longer make the losses in the two cases of the protocol vanish entirely. By choosing $g$ and $\varepsilon$ as functions of $b$ so as to balance these losses, we obtain the following theorem.

\begin{theorem}
    \label{thm:constant-v}
    There exists a positive constant $c$ such that the following holds. Let $U = \omega(n)$, and let $b \ge 2$ (where $b$ may depend on $n$). Consider a value-dynamic retrieval data structure that stores $b$-bit values for $n$ keys from $[U]$. Then the data structure uses at least
    $$nb + \left(\log e - c \sqrt{\frac{\log b}{b}}\right) n - o(n)$$
    bits of space in the worst case.
\end{theorem}

Note that, when $b = \omega(1)$, \cref{thm:constant-v} recovers \cref{thm:super-constant-v}. In comparison, the upper bound of \cite{kuszmaul2024space} achieves space $nb + (\log e - \Omega(2^{-b}))n + o(n)$ bits. Thus, the optimal redundancy is $(\log e - \delta_b) n \pm o(n)$ bits, where $\Omega(2^{-b}) \le \delta_b \le O(\sqrt{(\log b) / b})$. Determining the correct asymptotic behavior of $\delta_b$ remains an open question.

\paragraph{Choosing the parameters.} Before giving the proof, we explain where the bound $\sqrt{(\log b)/b}$ comes from. Recall that \cref{prop:general} gives $M \ge nb + \min\{\varepsilon s b,\ (1 - \varepsilon) n \log e\} - O(\frac{n \log g}{g}) - o(n)$, where $g$ is the number of batches and $s = \lfloor n/g \rfloor$. The first term in the minimum is $\varepsilon s b \approx \varepsilon n b / g$. For it not to be the bottleneck, we need $\varepsilon b / g \gtrsim \log e$, so we take $g \approx \varepsilon b$. The bound is then $n \log e$ minus two losses: a loss of $\varepsilon n \log e$ from the factor $1 - \varepsilon$, and a loss of $O(\frac{n \log g}{g}) = O(\frac{n \log(\varepsilon b)}{\varepsilon b})$ from the error term. The first loss increases with $\varepsilon$ and the second decreases with $\varepsilon$. Setting $\varepsilon \approx \frac{\log b}{\varepsilon b}$ balances them, which gives $\varepsilon \approx \sqrt{(\log b) / b}$ and a total loss of $O(n \sqrt{(\log b)/b})$.

\begin{proof}
    Let $b_0$ be a sufficiently large constant. If $b < b_0$, then $\sqrt{(\log b)/b} \ge 1/\sqrt{b_0}$, so for $c \ge \sqrt{b_0} \log e$, the claimed bound is at most $nb$, which holds trivially (Bob can recover all $n$ values by querying the data structure). So assume that $b \ge b_0$ and that $n$ is sufficiently large, and set
    $$\varepsilon = \sqrt{\frac{\log b}{b}} \quad\text{and}\quad g = \min\left(\left\lfloor \frac{\varepsilon b}{2 \log e} \right\rfloor, \left\lfloor \sqrt{n} \right\rfloor\right).$$
    Since $b_0$ and $n$ are sufficiently large, we have $\varepsilon \le 1/2$ and $g \ge 3$, so \cref{prop:general} applies. Since $g \le \frac{\varepsilon b}{2 \log e}$ and $s \ge n/(2g)$, the first term in the minimum is $\varepsilon s b \ge \frac{n}{2} \cdot \frac{\varepsilon b}{g} \ge n \log e$, so the minimum is $(1 - \varepsilon) n \log e$. For the error term, note that $g \ge \min(\frac{\varepsilon b}{4 \log e}, \frac{\sqrt{n}}{2})$, and that $(\log x)/x$ is decreasing for $x \ge 3$. Since $\varepsilon b = \sqrt{b \log b}$, this gives
    $$\frac{\log g}{g} = O\left(\frac{\log(\varepsilon b)}{\varepsilon b}\right) + O\left(\frac{\log n}{\sqrt{n}}\right) = O\left(\sqrt{\frac{\log b}{b}}\right) + o(1).$$
    So \cref{prop:general} gives
    $$M \ge nb + (1 - \varepsilon) n \log e - O\left(n \sqrt{\frac{\log b}{b}}\right) - o(n) = nb + \left(\log e - O\left(\sqrt{\frac{\log b}{b}}\right)\right) n - o(n),$$
    which proves the theorem for a sufficiently large constant $c$.
\end{proof}

\section{Lower Bound for Partially Value-Dynamic Retrieval}
\label{sec:partially-dynamic}

In this section, we prove our lower bound for partially value-dynamic retrieval. Throughout, we let $s = n/k$ denote the number of dynamic keys, and, as discussed in \cref{sec:prelim}, we assume without loss of generality that the data structure is deterministic.

\begin{theorem}
    \label{thm:partially-dynamic}
    There exist positive constants $c$ and $k_0$ such that the following holds. Let $k \ge k_0$, let $b \ge 1$, and let $U \ge 2n$. Consider a partially value-dynamic retrieval data structure that stores $b$-bit values for $n$ keys from $[U]$, of which $n/k$ are dynamic. Then the data structure uses at least
    $$nb + c \cdot \frac{n}{k} \log k$$
    bits of space in the worst case. In particular, if $k = \omega(1)$, then the redundancy is $\Omega((n/k) \log k)$.
\end{theorem}

\paragraph{Intuition.} As discussed in the introduction, the data structure must, in some sense, remember which $s$ of the $n$ keys are dynamic, which takes $\log \binom{n}{s} = \Theta(s \log k)$ bits. Accordingly, Alice's input will include the set $K$ of indices of the dynamic keys, and our goal is to show that a data structure with too little redundancy would let Alice compress her input.

The argument reuses the two ideas behind \cref{sec:warm-up,sec:super-constant-v}. There, Bob ruled out candidates for the real keys using the fact that real keys do not corrupt each other (Case~2), and Alice hid extra values in keys that can be updated without corrupting the real keys (Case~1). In the partially-dynamic setting, only the dynamic keys are guaranteed not to corrupt the other real keys (\cref{obs:legal-consistent}). So the static keys now play the role of the real keys that must be preserved, and the dynamic keys play the role of the keys that Alice wants to reveal cheaply. Call a set $A$ \emph{revealing} for $S_{\mathrm{dyn}}$ if updating $A$ preserves every static value but corrupts every dynamic key. Given such a set, Bob can tell which real keys are dynamic. We consider three cases.
\begin{itemize}
    \item If many sets are revealing for $S_{\mathrm{dyn}}$ (at least a $Q^{-10}$ fraction of all sets of $s$ keys, where $\log Q = \Theta(s \log k)$), then Alice can point to one of them using $O(\log Q)$ bits, which is much cheaper than sending $K$ directly.
    \item Otherwise, call a set a \emph{candidate dynamic set} if, like $S_{\mathrm{dyn}}$, it can be updated without corrupting any static key, and few sets are revealing for it. If there are few candidates, then, as in Case~2 of \cref{sec:super-constant-v}, Bob can use the keys that he already knows (here, the static keys) to rule out most possibilities for the remaining keys (here, the dynamic keys), and Alice saves $\log Q$ bits when sending them.
    \item If there are many candidates, then, as in Case~1 of \cref{sec:warm-up,sec:super-constant-v}, Alice uses them as extra storage. Unlike a deceptive set, however, a candidate set may corrupt the dynamic keys, and Bob does not know which candidate Alice used. Instead of telling him, Alice uses the fact that writing values into different candidates leads to distinct memory configurations, all of which agree on the static values, and she picks a configuration whose hash encodes an extra message. She then sends the dynamic values explicitly, at a cost of $sb$ bits. Since she can write almost $b$ bits into each of the $s$ keys of a candidate, and since there are about $Q^2$ candidates to choose from, she gains about $sb + 2 \log Q$ bits, for a net savings of about $2 \log Q$ bits.
\end{itemize}

\paragraph{The communication game.} Let $\gamma$ be a sufficiently small positive constant, and set
\begin{align*}
    Q\coloneqq k^{\gamma s},\qquad
    t\coloneqq \floor*{2 \log Q + s \log(2^b - 1)} - 2.
\end{align*}
Note that, since $s \ge 1$ and $k$ is sufficiently large, we have $Q \ge 2$, and hence $t \ge 0$. Intuitively, $\log Q = \gamma s \log k$ is (up to constant factors) the number of bits that we aim to save, and $t$ is the length of an extra message $z$ that Alice will try to hide inside the memory configuration in Case~3 below. The thresholds $Q^{-10}$ and $Q^{-1}$ that appear in the three cases are chosen so that each case saves $\Omega(\log Q)$ bits.

Consider the one-way communication game where Alice knows $n$ distinct, ordered keys $x_1,\dots,x_n\in[U]$, values $v_1,\dots,v_n\in[2^b]$, a set $K\subseteq[n]$ of size $s$, and an independent extra message $z\in[2^t]$, all chosen uniformly at random. Alice wishes to send all this information to Bob. The information-theoretic lower bound on the expected message length is
\begin{align*}
    H = \log \frac{U!}{(U-n)!} + nb + \log\binom{n}{s} + t.
\end{align*}

In the following, we describe a protocol using a partially value-dynamic retrieval data structure with at most $M$ bits of space. We will show that the expected message length is at most
\begin{align}
    \label{equ:partial-small-goal}
    H + M - nb - \Omega(s \log k).
\end{align}
By \cref{fact:source-coding}, this implies that $M \ge nb + \Omega(s \log k)$, which proves \cref{thm:partially-dynamic}.

\paragraph{Definitions.} Let $S=\{x_1,\dots,x_n\}$ be the real key set, with
\begin{align*}
    S_{\mathrm{dyn}}\coloneqq\{x_i:i\in K\},\qquad
    S_{\mathrm{stat}}\coloneqq\{x_i:i\notin K\}.
\end{align*}
Let $D$ denote the memory configuration of the data structure initialized with $(x_1,v_1),\dots,(x_n,v_n)$, where the keys in $S_{\mathrm{dyn}}$ are dynamic. We use the notions of consistency from \cref{def:consistency}, always with respect to $D$. Recall that a set $A$ is consistent w.r.t.\ a key $x$ if $A$ is internally consistent (values written to the keys of $A$ can be read back) and no sequence of updates to the keys of $A$ changes the answer to $\textsc{Query}(x)$.

For sets $A, B \subseteq [U]$, we say that \defn{$A$ is revealing for $B$}, if $A$ has size $s$, is internally consistent, is consistent w.r.t.\ every key in $S_{\mathrm{stat}}$, and is inconsistent w.r.t.\ every key in $B$. Define
\begin{align*}
    p(B)\coloneqq
    \Pr_{A}
    [A\text{ is revealing for }B],
\end{align*}
where $A$ is a uniformly random subset of $[U]$ of size $s$. The name reflects the fact that a set $A$ that is revealing for $S_{\mathrm{dyn}}$ would reveal to Bob which keys are dynamic, since, given $A$, the dynamic keys are exactly the real keys that $A$ can corrupt. Thus $p(S_{\mathrm{dyn}})$ measures how easy it is to find a set that reveals $S_{\mathrm{dyn}}$ by random sampling.

We say that $B$ is a \defn{candidate dynamic set}, if
\begin{itemize}
    \item $|B|=s$ and $B\cap S_{\mathrm{stat}}=\emptyset$;
    \item $B$ is internally consistent, and is consistent w.r.t.\ every key in $S_{\mathrm{stat}}$;
    \item $p(B)\le Q^{-10}$.
\end{itemize}
In other words, a candidate dynamic set is a set that could plausibly be the set of dynamic keys, from the point of view of someone who knows only $D$ and $S_{\mathrm{stat}}$. It is disjoint from the static keys, it can be updated without corrupting them, and few sets are revealing for it. Let $C$ denote the collection of all candidate dynamic sets. Given $D$ and $S_{\mathrm{stat}}$, Bob can compute $C$ without knowing $S_{\mathrm{dyn}}$. Our protocol considers the following three cases.

\paragraph{Case 1: Many sets are revealing for $S_{\mathrm{dyn}}$.} Suppose that $p(S_{\mathrm{dyn}})>Q^{-10}$. Using public randomness, sample an infinite sequence $A_1, A_2, \ldots$ of independent uniformly random subsets of $[U]$ of size $s$, and let $\iota$ be the smallest index such that $A_\iota$ is revealing for $S_{\mathrm{dyn}}$. Alice sends the following message to Bob:
\begin{itemize}
    \item Two bits indicating that we are in case 1.
    \item The data structure $D$, using $M$ bits.
    \item The keys $x_1,\dots,x_n$, using $\lceil \log \frac{U!}{(U-n)!} \rceil$ bits.
    \item The index $\iota$, using $O(1 + \log \iota)$ bits.
    \item The extra message $z$, using $t$ bits.
\end{itemize}

\paragraph{Correctness.} Bob recovers the values $v_1,\dots,v_n$ by querying $D$. He also recovers $A_\iota$ from its index. By the definition of a revealing set, $A_\iota$ is consistent w.r.t.\ every static key, and inconsistent w.r.t.\ every dynamic key. Thus Bob recovers $K = \{i \mid A_\iota \text{ is inconsistent w.r.t.\ } x_i\}$.

\paragraph{Length of message.} The index $\iota$ is a geometric random variable with success probability $p(S_{\mathrm{dyn}}) > Q^{-10}$ (each $A_i$ is revealing for $S_{\mathrm{dyn}}$ independently with probability $p(S_{\mathrm{dyn}})$), so $\E[\iota] < Q^{10}$, and by Jensen's inequality, the expected length of its encoding is $O(1 + \log \E[\iota]) = O(\log Q)$ (recall that $Q \ge 2$). The total expected message length is thus at most
\begin{align*}
    M + \log \frac{U!}{(U-n)!} + t + O(\log Q) = H + M - nb - \left(\log \binom{n}{s} - O(\log Q)\right).
\end{align*}
Since $\log \binom{n}{s} \ge s \log(n/s) = s \log k$, and since $O(\log Q) = O(\gamma s \log k) \le \frac{1}{2} s \log k$ (as $\gamma$ is sufficiently small), the quantity in parentheses is at least $\frac{1}{2} s \log k$, which gives \eqref{equ:partial-small-goal}.

\paragraph{Case 2: Few sets are revealing, and there are few candidate dynamic sets.} Now suppose that $p(S_{\mathrm{dyn}})\le Q^{-10}$ and
\begin{align*}
    |C|\le Q^{-1}\binom{U}{s}.
\end{align*}
The real dynamic set $S_{\mathrm{dyn}}$ belongs to $C$. Indeed, it is disjoint from $S_{\mathrm{stat}}$; since updates to dynamic keys are legal, it is internally consistent and preserves every static value; and $p(S_{\mathrm{dyn}}) \le Q^{-10}$ by assumption. Alice sends the following message to Bob:
\begin{itemize}
    \item Two bits indicating that we are in case 2.
    \item The data structure $D$, using $M$ bits.
    \item The set $K$, using $\lceil \log\binom{n}{s} \rceil$ bits.
    \item The keys $x_i$ for $i\notin K$, in their original order, using $\lceil \log \frac{U!}{(U-n+s)!} \rceil$ bits.
    \item The set $S_{\mathrm{dyn}}$, encoded as a member of $C$, using $\lceil \log |C| \rceil$ bits, followed by the order in which its keys appear among the positions in $K$, using $\lceil \log (s!) \rceil$ bits.
    \item The extra message $z$, using $t$ bits.
\end{itemize}

\paragraph{Correctness.} After receiving the static keys, Bob can compute $C$ and decode $S_{\mathrm{dyn}}$. Its ordering and $K$ specify the original sequence $x_1,\dots,x_n$. Bob then queries $D$ to recover $v_1,\dots,v_n$.

\paragraph{Length of message.} Sending the dynamic keys and their ordering costs at most
\begin{align*}
    \log|C|+\log(s!) + 2
    &\le \log\binom{U}{s}+\log(s!)-\log Q + 2 = \log \frac{U!}{(U-s)!} - \log Q + 2.
\end{align*}
We will use the following estimate, which holds for $s \le n \le U/2$:
\begin{equation}
    \log \frac{U!}{(U-s)!} + \log \frac{U!}{(U-n+s)!} - \log \frac{U!}{(U-n)!} = \sum_{i = 0}^{s - 1} \log \frac{U - i}{U - n + s - i} \le s \log \frac{U}{U - n} \le s.
    \label{eq:ff-split}
\end{equation}
In words, encoding the static keys and the dynamic keys separately, as if they did not have to be distinct from each other, costs at most $s$ extra bits. Thus the total message length is at most
\begin{align*}
    & M + \log \binom{n}{s} + \log \frac{U!}{(U-n+s)!} + \log \frac{U!}{(U-s)!} - \log Q + t + O(1) \\
    \le {} & M + \log \binom{n}{s} + \log \frac{U!}{(U-n)!} + O(s) - \log Q + t + O(1) \\
    = {} & H + M - nb - \left(\gamma s \log k - O(s)\right),
\end{align*}
where the inequality uses \eqref{eq:ff-split}. Since $k$ is sufficiently large, this gives \eqref{equ:partial-small-goal}.

\paragraph{Case 3: Few sets are revealing, and there are many candidate dynamic sets.} Finally, suppose that
\begin{align*}
    |C|>Q^{-1}\binom{U}{s}.
\end{align*}
(This case does not use the assumption that $p(S_{\mathrm{dyn}}) \le Q^{-10}$.) The idea is to find many candidate dynamic sets whose updates lead to distinct memory configurations, all preserving the static values.

\begin{claim}
    \label{clm:non-revealing-family}
    There exist $\ell = \lceil Q^2/2 \rceil$ distinct candidate dynamic sets $L_1, \ldots, L_\ell \in C$ such that $L_i$ is not revealing for $L_j$ whenever $i \neq j$.
\end{claim}
Intuitively, candidate dynamic sets make up more than a $1/Q$ fraction of all $s$-subsets of $[U]$, while each candidate is revealed by at most a $Q^{-10}$ fraction of them. So, if we sample about $Q^3$ random sets, then about $Q^2$ of them will be candidates, and it is unlikely that any one of them is revealing for another.

\begin{proof}
    We use the probabilistic method. Let $N = \lceil Q^3 \rceil$, and sample $N$ independent uniformly random sets $T_1,\dots,T_{N}$ of size $s$. Let $J = \{i \mid T_i \in C\}$. Then $\E[|J|] = N |C| / \binom{U}{s} > Q^2$, so by a (multiplicative) Chernoff bound for the sum of the independent indicators $\ind[T_i \in C]$, $\Pr[|J| < Q^2/2] \le \Pr[|J| < \E[|J|]/2] \le e^{-\E[|J|]/8} \le e^{-Q^2/8}$.

    Call an ordered pair $(i, j)$ of distinct indices \emph{bad} if $T_j \in C$ and $T_i$ is revealing for $T_j$. Since $T_i$ and $T_j$ are independent, and since every $B \in C$ satisfies $p(B) \le Q^{-10}$, each pair is bad with probability at most $Q^{-10}$. There are at most $N^2 \le 4Q^6$ pairs, so the probability that any pair is bad is at most $4Q^{-4}$.

    Since $Q \ge 2$, we have $e^{-Q^2/8} + 4Q^{-4} < 1$, so with positive probability we have both $|J| \ge Q^2/2$ and no bad pairs. Fix such an outcome. The sets $T_i$ for $i \in J$ are pairwise non-revealing (in the sense of the claim). They are also distinct. To see this, suppose that $T_i = T_j$ for some distinct $i, j \in J$. Then, since $T_j \in C$ is internally consistent and consistent w.r.t.\ $S_{\mathrm{stat}}$, and is inconsistent w.r.t.\ each of its own keys (as is every set, see the remarks after \cref{def:consistency}), $T_i$ would be revealing for $T_j$, and $(i, j)$ would be bad. Taking $L_1, \ldots, L_\ell$ to be any $\ell$ of the sets $T_i$, $i \in J$, completes the proof.
\end{proof}

Alice fixes the first such family $L_1, \ldots, L_\ell$ (under some canonical order). Note that Alice's choice is a deterministic function of $D$ and $S_{\mathrm{stat}}$. Bob will not need to know it.

\paragraph{Distinct memory configurations.} For each $i \in [\ell]$, let $W_i$ denote the set of assignments $w$ of values in $[2^b]$ to the keys of $L_i$ such that every key $y \in L_i$ is assigned a value $w_y$ that differs from its answer on $D$. Thus $|W_i| = (2^b - 1)^s$. For $w \in W_i$, let $D_{i, w}$ be the data structure obtained from $D$ by performing $\textsc{Update}(y, w_y)$ for every $y \in L_i$, in a fixed canonical order. Since $L_i$ is internally consistent, every key $y \in L_i$ is answered with $w_y$ on $D_{i, w}$, which differs from its answer on $D$. Since $L_i$ is consistent w.r.t.\ $S_{\mathrm{stat}}$, all static values are unchanged.

We claim that the $N' = \ell \cdot (2^b - 1)^s$ configurations $D_{i, w}$ are distinct. For a fixed $i$, two different assignments $w \ne w'$ in $W_i$ differ on some key of $L_i$, so $D_{i, w} \ne D_{i, w'}$. Suppose that $D_{i, w} = D_{j, w'}$ for some $i \ne j$. Then the updates to $L_i$ change the answer of every key in $L_j$, because those answers are changed on $D_{j, w'}$. Thus $L_i$ is inconsistent w.r.t.\ every key in $L_j$, and so $L_i$ is revealing for $L_j$, a contradiction. (When $b = 1$, each $W_i$ consists of a single assignment, which flips every value.)

\paragraph{Encoding the extra message.} Alice now has $N' \ge \frac{Q^2}{2} (2^b - 1)^s$ different memory configurations to choose from, all of which agree with $D$ on the static values. By the choice of $t$, this is at least twice the number $2^t$ of possible extra messages, so she can typically find a configuration whose hash is $z$, and Bob can then read $z$ off by hashing the configuration. The price is that the dynamic values may be destroyed in the chosen configuration, so Alice sends them explicitly, at a cost of $sb$ bits. This is paid for by the term $s \log(2^b - 1)$ in $t$, which reflects that Alice can write almost $b$ bits of her choice into each key of a candidate set. Using public randomness, sample an infinite sequence $h_1, h_2, \ldots$ of independent fully random hash functions from $M$-bit strings to $[2^t]$. For any fixed $j$ and any fixed extra message $z$, the probability that none of the $N'$ configurations $D_{i, w}$ hashes to $z$ under $h_j$ is at most
\begin{align*}
    (1-2^{-t})^{N'}
    \le \exp\bk*{-\frac{N'}{2^t}}
    \le e^{-2} \le \frac{1}{2},
\end{align*}
where the second inequality uses $2^t \le \frac{Q^2}{4} (2^b - 1)^s \le N'/2$. (Here we use that the configurations $D_{i, w}$ are distinct, so that their hash values under $h_j$ are independent and uniformly random.) Let $\lambda$ be the smallest index such that $h_\lambda(D_{i, w}) = z$ for some $i$ and $w \in W_i$, and let $D'$ be the first such configuration (under some canonical order). Then $\E[\lambda] \le 2$. Alice sends the following message to Bob:
\begin{itemize}
    \item Two bits indicating that we are in case 3.
    \item The data structure $D'$, using $M$ bits.
    \item The keys $x_1,\dots,x_n$, using $\lceil \log \frac{U!}{(U-n)!} \rceil$ bits.
    \item The set $K$, using $\lceil \log\binom{n}{s} \rceil$ bits.
    \item The dynamic values $v_i$ for $i\in K$, using $s\cdot b$ bits.
    \item The index $\lambda$, using $O(1 + \log \lambda)$ bits.
\end{itemize}

\paragraph{Correctness.} Bob queries $D'$ to recover the static values, which are unchanged. The dynamic values are sent explicitly. Finally, Bob recovers the extra message as $z=h_\lambda(D')$. He does not need to recover the candidate set used to obtain $D'$.

\paragraph{Length of message.} The encoding of $\lambda$ has expected length $O(1 + \log \E[\lambda]) = O(1)$, by Jensen's inequality and $\E[\lambda] \le 2$. So the total expected message length is at most
\begin{align*}
    M+\log \frac{U!}{(U-n)!} +\log\binom{n}{s}
    +s\cdot b+O(1) = H + M - nb - \left(t - s \cdot b - O(1)\right).
\end{align*}
Since $2^b - 1 \ge 2^{b - 1}$, we have $t \ge 2 \log Q + s(b - 1) - 3$, so the quantity in parentheses is at least $2 \log Q - s - O(1) = 2 \gamma s \log k - s - O(1)$. Since $k$ is sufficiently large (as a function of $\gamma$), this gives \eqref{equ:partial-small-goal}.

This completes the proof of \cref{thm:partially-dynamic}.

\section*{Funding Acknowledgments}

Yang S. Hu, William Kuszmaul, Jingxun Liang, and Renfei Zhou were partially supported by NSF grant CNS-2504471, by NSF CAREER award CCF-2542165, and by a Jane Street Research Grant. Renfei Zhou was partially supported by the Jane Street Graduate Research Fellowship and the MongoDB PhD Fellowship.

\section*{AI Disclosure}

All of the technical content of this paper, including its results and proofs, was developed by the authors in 2025, before the emergence of AI systems capable of contributing to research-level mathematics. Such systems were, however, used in preparing the manuscript, for example to draft the exposition and check for errors. The authors take full responsibility for the content and correctness of the final manuscript.

\bibliographystyle{alpha}
\bibliography{reference.bib}

\newcommand{\etalchar}[1]{$^{#1}$}
\begin{thebibliography}{DHSSW22}

\bibitem[ADR09]{aumuller2009experimental}
Martin Aum{\"u}ller, Martin Dietzfelbinger, and Michael Rink.
\newblock Experimental variations of a theoretically good retrieval data
  structure.
\newblock In {\em Proc. 17th European Symposium on Algorithms (ESA)}, pages
  742--751, 2009.

\bibitem[AFK23]{assadi2023tight}
Sepehr Assadi, Mart{\'i}n {Farach-Colton}, and William Kuszmaul.
\newblock Tight bounds for monotone minimal perfect hashing.
\newblock In {\em Proc. 34th ACM-SIAM Symposium on Discrete Algorithms (SODA)},
  pages 456--476, 2023.

\bibitem[ANS10]{arbitman2010backyard}
Yuriy Arbitman, Moni Naor, and Gil Segev.
\newblock Backyard cuckoo hashing: constant worst-case operations with a
  succinct representation.
\newblock In {\em Proc. 51st IEEE Symposium on Foundations of Computer Science
  (FOCS)}, pages 787--796, 2010.

\bibitem[BBD09]{belazzougui2009hash}
Djamal Belazzougui, Fabiano~C. Botelho, and Martin Dietzfelbinger.
\newblock Hash, displace, and compress.
\newblock In {\em Proc. 17th European Symposium on Algorithms (ESA)}, pages
  682--693, 2009.

\bibitem[BBPV09]{belazzougui2009monotone}
Djamal Belazzougui, Paolo Boldi, Rasmus Pagh, and Sebastiano Vigna.
\newblock Monotone minimal perfect hashing: searching a sorted table with
  {$O(1)$} accesses.
\newblock In {\em Proc. 20th ACM-SIAM Symposium on Discrete Algorithms (SODA)},
  pages 785--794, 2009.

\bibitem[BE20]{BE20}
Ioana~O. Bercea and Guy Even.
\newblock A dynamic space-efficient filter with constant time operations.
\newblock In {\em Proc. 17th Scandinavian Symposium and Workshops on Algorithm
  Theory (SWAT)}, pages 11:1--11:17, 2020.

\bibitem[BFK{\etalchar{+}}22]{bender2022optimal}
Michael~A. Bender, Mart{\'i}n {Farach-Colton}, John Kuszmaul, William Kuszmaul,
  and Mingmou Liu.
\newblock On the optimal time/space tradeoff for hash tables.
\newblock In {\em Proc. 54th ACM Symposium on Theory of Computing (STOC)},
  pages 1284--1297, 2022.

\bibitem[CFG{\etalchar{+}}78]{carter1978exact}
Larry Carter, Robert~W. Floyd, John Gill, George Markowsky, and Mark~N. Wegman.
\newblock Exact and approximate membership testers.
\newblock In {\em Proc. 10th ACM Symposium on Theory of Computing (STOC)},
  pages 59--65, 1978.

\bibitem[CKRT04]{chazelle2004bloomier}
Bernard Chazelle, Joe Kilian, Ronitt Rubinfeld, and Ayellet Tal.
\newblock The {Bloomier} filter: an efficient data structure for static support
  lookup tables.
\newblock In {\em Proc. 15th ACM-SIAM Symposium on Discrete Algorithms (SODA)},
  pages 30--39, 2004.

\bibitem[CT06]{CoverThomas06}
Thomas~M. Cover and Joy~A. Thomas.
\newblock {\em Elements of Information Theory}.
\newblock Wiley, 2nd edition, 2006.

\bibitem[DHSSW22]{dillinger2022fast}
Peter~C. Dillinger, Lorenz H{\"u}bschle-Schneider, Peter Sanders, and Stefan
  Walzer.
\newblock Fast succinct retrieval and approximate membership using ribbon.
\newblock In {\em Proc. 20th International Symposium on Experimental Algorithms
  (SEA)}, pages 4:1--4:20, 2022.

\bibitem[Die07]{dietzfelbinger2007design}
Martin Dietzfelbinger.
\newblock Design strategies for minimal perfect hash functions.
\newblock In {\em Proc. 4th International Symposium on Stochastic Algorithms:
  Foundations and Applications (SAGA)}, pages 2--17, 2007.

\bibitem[DMPP06]{demaine2006dictionariis}
Erik~D. Demaine, Friedhelm {Meyer auf der Heide}, Rasmus Pagh, and Mihai P{\v
  a}tra{\c s}cu.
\newblock De dictionariis dynamicis pauco spatio utentibus.
\newblock In {\em Proc. 7th Latin American Symposium on Theoretical Informatics
  (LATIN)}, pages 349--361, 2006.

\bibitem[DP08]{dietzfelbinger2008succinct}
Martin Dietzfelbinger and Rasmus Pagh.
\newblock Succinct data structures for retrieval and approximate membership.
\newblock In {\em Proc. 35th International Colloquium on Automata, Languages
  and Programming (ICALP)}, pages 385--396, 2008.

\bibitem[DW19]{dietzfelbinger2019constanttime}
Martin Dietzfelbinger and Stefan Walzer.
\newblock Constant-time retrieval with {$O(\log m)$} extra bits.
\newblock In {\em Proc. 36th International Symposium on Theoretical Aspects of
  Computer Science (STACS)}, pages 24:1--24:16, 2019.

\bibitem[DW21]{dillinger2021ribbon}
Peter~C. Dillinger and Stefan Walzer.
\newblock Ribbon filter: practically smaller than {Bloom} and {Xor}.
\newblock arXiv:2103.02515, 2021.

\bibitem[EMV20]{esposito2020recsplit}
Emmanuel Esposito, Thomas {Mueller Graf}, and Sebastiano Vigna.
\newblock {RecSplit}: minimal perfect hashing via recursive splitting.
\newblock In {\em Proc. 22nd Workshop on Algorithm Engineering and Experiments
  (ALENEX)}, pages 175--185, 2020.

\bibitem[FK84]{fredman1984size}
Michael~L. Fredman and J{\'a}nos Koml{\'o}s.
\newblock On the size of separating systems and families of perfect hash
  functions.
\newblock {\em SIAM Journal on Algebraic Discrete Methods}, 5(1):61--68, 1984.

\bibitem[FKS84]{fredman1984storing}
Michael~L. Fredman, J{\'a}nos Koml{\'o}s, and Endre Szemer{\'e}di.
\newblock Storing a sparse table with {$O(1)$} worst case access time.
\newblock {\em Journal of the ACM}, 31(3):538--544, 1984.

\bibitem[HKL{\etalchar{+}}25]{hu2025static}
Yang Hu, William Kuszmaul, Jingxun Liang, Huacheng Yu, Junkai Zhang, and Renfei
  Zhou.
\newblock Static retrieval revisited: to optimality and beyond.
\newblock In {\em Proc. 66th IEEE Symposium on Foundations of Computer Science
  (FOCS)}, pages 2392--2409, 2025.

\bibitem[HT01]{HT01}
Torben Hagerup and Torsten Tholey.
\newblock Efficient minimal perfect hashing in nearly minimal space.
\newblock In {\em Proc. 18th International Symposium on Theoretical Aspects of
  Computer Science (STACS)}, pages 317--326, 2001.

\bibitem[KLZ25]{kuszmaul2025fingerprint}
William Kuszmaul, Jingxun Liang, and Renfei Zhou.
\newblock Fingerprint filters are optimal.
\newblock In {\em Proc. 66th IEEE Symposium on Foundations of Computer Science
  (FOCS)}, pages 1059--1073, 2025.

\bibitem[Kos24]{kosolobov2024simplified}
Dmitry Kosolobov.
\newblock Simplified tight bounds for monotone minimal perfect hashing.
\newblock In {\em Proc. 35th Symposium on Combinatorial Pattern Matching
  (CPM)}, pages 19:1--19:13, 2024.

\bibitem[KPX{\etalchar{+}}25]{kuszmaul2025tight}
William Kuszmaul, Aaron Putterman, Tingqiang Xu, Hangrui Zhou, and Renfei Zhou.
\newblock Tight bounds and phase transitions for incremental and dynamic
  retrieval.
\newblock In {\em Proc. 36th ACM-SIAM Symposium on Discrete Algorithms (SODA)},
  pages 3974--3997, 2025.

\bibitem[KPX{\etalchar{+}}26]{kuszmaul2026resizable}
William Kuszmaul, Aaron Putterman, Tingqiang Xu, Hangrui Zhou, and Renfei Zhou.
\newblock Resizable retrieval.
\newblock arXiv:2606.15944, 2026.

\bibitem[KW24]{kuszmaul2024space}
William Kuszmaul and Stefan Walzer.
\newblock Space lower bounds for dynamic filters and value-dynamic retrieval.
\newblock In {\em Proc. 56th ACM Symposium on Theory of Computing (STOC)},
  pages 1153--1164, 2024.

\bibitem[LMP{\etalchar{+}}26]{lehmann2026modern}
Hans-Peter Lehmann, Thomas Mueller, Rasmus Pagh, Giulio~Ermanno Pibiri, Peter
  Sanders, Sebastiano Vigna, and Stefan Walzer.
\newblock Modern minimal perfect hashing: a survey.
\newblock {\em ACM Computing Surveys}, 58(10):1--36, 2026.

\bibitem[LP13]{lovett2013space}
Shachar Lovett and Ely Porat.
\newblock A space lower bound for dynamic approximate membership data
  structures.
\newblock {\em SIAM Journal on Computing}, 42(6):2182--2196, 2013.

\bibitem[LSW24]{lehmann2024shockhash}
Hans-Peter Lehmann, Peter Sanders, and Stefan Walzer.
\newblock {ShockHash}: towards optimal-space minimal perfect hashing beyond
  brute-force.
\newblock In {\em Proc. 26th Symposium on Algorithm Engineering and Experiments
  (ALENEX)}, pages 194--206, 2024.

\bibitem[LYY20]{liu2020succinct}
Mingmou Liu, Yitong Yin, and Huacheng Yu.
\newblock Succinct filters for sets of unknown sizes.
\newblock In {\em Proc. 47th International Colloquium on Automata, Languages
  and Programming (ICALP)}, pages 79:1--79:19, 2020.

\bibitem[Meh82]{mehlhorn1982program}
Kurt Mehlhorn.
\newblock On the program size of perfect and universal hash functions.
\newblock In {\em Proc. 23rd IEEE Symposium on Foundations of Computer Science
  (FOCS)}, pages 170--175, 1982.

\bibitem[MPP05]{mortensen2005dynamic}
Christian~Worm Mortensen, Rasmus Pagh, and Mihai P{\v a}tra{\c s}cu.
\newblock On dynamic range reporting in one dimension.
\newblock In {\em Proc. 37th ACM Symposium on Theory of Computing (STOC)},
  pages 104--111, 2005.

\bibitem[Por09]{porat2009optimal}
Ely Porat.
\newblock An optimal {Bloom} filter replacement based on matrix solving.
\newblock In {\em Proc. 4th International Computer Science Symposium in Russia
  (CSR)}, pages 263--273, 2009.

\bibitem[PPR05]{pagh2005optimal}
Anna Pagh, Rasmus Pagh, and S.~Srinivasa Rao.
\newblock An optimal {Bloom} filter replacement.
\newblock In {\em Proc. 16th ACM-SIAM Symposium on Discrete Algorithms (SODA)},
  pages 823--829, 2005.

\bibitem[PSW13]{pagh2013how}
Rasmus Pagh, Gil Segev, and Udi Wieder.
\newblock How to approximate a set without knowing its size in advance.
\newblock In {\em Proc. 54th IEEE Symposium on Foundations of Computer Science
  (FOCS)}, pages 80--89, 2013.

\bibitem[SS90]{SS90}
Jeanette~P. Schmidt and Alan Siegel.
\newblock The spatial complexity of oblivious $k$-probe hash functions.
\newblock {\em SIAM Journal on Computing}, 19(5):775--786, 1990.

\end{thebibliography}

\appendix
\crefalias{section}{appendix}
\crefalias{subsection}{appendix}

\section{Upper Bound for Partially Value-Dynamic Retrieval}
\label{sec:partially-dynamic-upper}

In this appendix, we show that the lower bound in \cref{thm:partially-dynamic} is tight up to a constant factor in the redundancy when $U \le 2^{\poly(n)}$. As in \cref{sec:partially-dynamic}, we let $s = n/k$ denote the number of dynamic keys. In \cref{sec:partially-dynamic-upper-space}, we give a construction that focuses only on space, matching the information-theoretic nature of our lower bounds. In \cref{sec:partially-dynamic-tradeoff}, we combine this construction with the results of Hu et al.\ \cite{hu2025static} on static retrieval to obtain tight bounds on the tradeoff between space and query time.

\subsection{An Information-Theoretic Upper Bound}
\label{sec:partially-dynamic-upper-space}

In this subsection, we make no attempt to make the operations time-efficient.

\begin{proposition}
    \label{prop:partially-dynamic-upper}
    Let $k \ge 2$. There is a partially value-dynamic retrieval data structure that stores $b$-bit values for $n$ keys from $[U]$, of which $n/k$ are dynamic, using
    $$nb + \frac{n}{k} \log k + O\left(\frac{n}{k} + \log n + \log \log U\right)$$
    bits of space.
\end{proposition}

\paragraph{Intuition.}
Our lower bound says that the data structure must, in some sense, remember which keys are dynamic. The construction does this as cheaply as possible. Storing one bit per key to indicate whether the key is dynamic would cost $n$ bits, which is too much. Instead, we first use a filter with false-positive rate $1/k$ to narrow the candidates for the dynamic keys down to a set of $O(s)$ keys, and only then store one bit per candidate. (Recall that a filter answers approximate membership queries: it accepts every key in a given set, and accepts any other key with probability at most its false-positive rate.) The filter costs $\log k + O(1)$ bits per dynamic key, and the dynamic values themselves are stored using minimal perfect hashing, as in the classical solution for value-dynamic retrieval. The static values are stored in a static retrieval data structure, which has almost no redundancy.

\paragraph{Fixing the randomness.}
Each component of our data structure is constructed by choosing a random function and checking that it satisfies some property. To obtain a deterministic data structure with a worst-case space bound, we will use the following standard application of the probabilistic method.

\begin{lemma}
    \label{lem:fix-randomness}
    Let $\mathcal{I}$ be a nonempty finite set (of \emph{instances}), and let $\mathcal{D}$ be a probability distribution over functions $h$ with some fixed domain and codomain. Suppose that, for each instance $I \in \mathcal{I}$, a random $h \sim \mathcal{D}$ is \emph{good} for $I$ with probability at least $p > 0$. Then there exist functions $h_1, \ldots, h_T$, where $T = \lfloor \ln |\mathcal{I}| / p \rfloor + 1$, such that every $I \in \mathcal{I}$ has at least one good function among $h_1, \ldots, h_T$. In particular, given $I$, the index of such a function can be written using
    $$\lceil \log T \rceil \le \log \frac{1}{p} + \log\left(\ln |\mathcal{I}| + 1\right) + 1$$
    bits.
\end{lemma}

\begin{proof}
    Draw $h_1, \ldots, h_T$ independently from $\mathcal{D}$. For each fixed $I \in \mathcal{I}$, the probability that none of $h_1, \ldots, h_T$ is good for $I$ is at most $(1 - p)^T \le e^{-pT} < 1/|\mathcal{I}|$. By a union bound over $\mathcal{I}$, there is a positive probability that every $I \in \mathcal{I}$ has a good function among $h_1, \ldots, h_T$, so some fixed choice of $h_1, \ldots, h_T$ has this property. The bound on $\lceil \log T \rceil$ follows from $T \le \ln |\mathcal{I}| / p + 1 \le (\ln |\mathcal{I}| + 1)/p$, where the final step uses $p \le 1$.
\end{proof}

The list $h_1, \ldots, h_T$ depends only on the parameters of the problem (and not on the input), so it can be hard-wired into the data structure's algorithms. This is consistent with the computational model of \cref{sec:prelim}, which requires only that the operations not depend on additional information about $S$.

\paragraph{Building blocks.}
We will use \cref{lem:fix-randomness} to construct the following two building blocks. Throughout, we identify $r$-bit strings with vectors in $\mathbb{F}_2^r$.

\begin{lemma}[Static retrieval]
    \label{lem:static-retrieval}
    Let $1 \le m \le U$ and $r \ge 0$. There is a static retrieval data structure that, given any set $X \subseteq [U]$ of $m$ keys and any function $f : X \to \mathbb{F}_2^r$, uses $mr + O(\log m + \log \log U)$ bits of space, and correctly answers queries to every key in $X$.
\end{lemma}

\begin{proof}
    Call a function $h : [U] \to \mathbb{F}_2^m$ good for $X$ if the $m$ vectors $\{h(x) \mid x \in X\}$ are linearly independent. If $h$ is uniformly random, then these $m$ vectors are independent and uniformly random, so, revealing them one at a time, the probability that each one lies outside the span of the previous ones is
    $$\prod_{i = 0}^{m - 1} \left(1 - \frac{2^i}{2^m}\right) = \prod_{j = 1}^{m} \left(1 - 2^{-j}\right) \ge \frac{1}{2} \left(1 - \sum_{j = 2}^{\infty} 2^{-j}\right) = \frac{1}{4}.$$
    Applying \cref{lem:fix-randomness} with $\mathcal{I}$ being the set of all $X \subseteq [U]$ of size $m$ (so $\ln |\mathcal{I}| \le m \ln U$) and $p = 1/4$, we obtain a fixed list of functions such that the index $i$ of a function $h_i$ that is good for $X$ can be written using $O(\log m + \log \log U)$ bits.

    The data structure stores $i$, along with vectors $z_1, \ldots, z_r \in \mathbb{F}_2^m$ such that, for every $x \in X$ and every $t \in [r]$, the inner product $\langle h_i(x), z_t \rangle$ equals the $t$-th bit of $f(x)$. Such vectors exist because $\{h_i(x) \mid x \in X\}$ is a basis of $\mathbb{F}_2^m$. To answer a query to $x$, the data structure returns the vector $(\langle h_i(x), z_1 \rangle, \ldots, \langle h_i(x), z_r \rangle)$. The total space is $mr + O(\log m + \log \log U)$ bits. (In other words, each $z_t$ is the solution of an $m \times m$ linear system over $\mathbb{F}_2$, with one equation per key, and the system is solvable because $h_i$ maps the $m$ keys to linearly independent vectors.)
\end{proof}

\begin{lemma}[Minimal perfect hashing]
    \label{lem:mphf}
    Let $1 \le m \le U$. Given any set $X \subseteq [U]$ of $m$ keys, one can store a function $h : [U] \to [m]$ that is injective on $X$ using $m \log e + O(\log m + \log \log U)$ bits.
\end{lemma}

\begin{proof}
    A uniformly random function $h : [U] \to [m]$ is injective on $X$ with probability $m! / m^m \ge e^{-m}$, where the inequality follows from $e^m = \sum_{j \ge 0} m^j / j! \ge m^m / m!$. Applying \cref{lem:fix-randomness} with $\mathcal{I}$ being the set of all $X \subseteq [U]$ of size $m$ and $p = e^{-m}$, we can store the index of a function that is injective on $X$ using $m \log e + \log(m \ln U + 1) + 1$ bits.
\end{proof}

\paragraph{The data structure.}
Let $\ell = \lceil \log k \rceil$, so that $2^{-\ell} \le 1/k$ and $\ell \le \log k + 1$. The data structure consists of the following components.
\begin{itemize}
    \item \emph{A filter for the dynamic keys.} Let $g : [U] \to \mathbb{F}_2^\ell$ be a function, which we call the \emph{fingerprint} function, and which will be chosen below. The data structure stores a static retrieval data structure $R_{\mathrm{fp}}$ (\cref{lem:static-retrieval}) that maps each dynamic key $x$ to its fingerprint $g(x)$. We say that a key $y$ \emph{passes} the filter if $R_{\mathrm{fp}}(y) = g(y)$. Every dynamic key passes the filter. Let $P \subseteq S_{\mathrm{stat}}$ denote the set of static keys that pass the filter.

    We now choose $g$. Note that the contents of $R_{\mathrm{fp}}$, and hence the answer $R_{\mathrm{fp}}(y)$ for each key $y$, are determined by $S_{\mathrm{dyn}}$ and by the restriction of $g$ to $S_{\mathrm{dyn}}$. If $g$ is uniformly random, then, for each $y \in S_{\mathrm{stat}}$, the fingerprint $g(y)$ is uniformly random and independent of $R_{\mathrm{fp}}(y)$, so $y$ passes the filter with probability $2^{-\ell} \le 1/k$. Thus $\E[|P|] \le (n - s)/k \le s$, and, by Markov's inequality, $|P| \le 2s$ with probability at least $1/2$. Applying \cref{lem:fix-randomness} with $\mathcal{I}$ being the set of all pairs $(S_{\mathrm{dyn}}, S_{\mathrm{stat}})$ (so $\ln |\mathcal{I}| \le n \ln (2U)$) and $p = 1/2$, the data structure can store the index of a fingerprint function $g$ for which $|P| \le 2s$, using $O(\log n + \log \log U)$ bits. Including this index, the filter uses $s\ell + O(\log n + \log \log U) \le s \log k + s + O(\log n + \log \log U)$ bits.
    \item \emph{A type bit for each key that passes the filter.} A static retrieval data structure $R_{\mathrm{type}}$ (\cref{lem:static-retrieval}) that maps each key in $S_{\mathrm{dyn}} \cup P$ to a single bit indicating whether that key is dynamic. The data structure also stores $|S_{\mathrm{dyn}} \cup P| \le 3s$, using $O(\log n)$ bits, so that the size of $R_{\mathrm{type}}$ is known. In total, this uses at most $3s + O(\log n + \log \log U)$ bits.
    \item \emph{The values of the static keys.} A static retrieval data structure $R_{\mathrm{stat}}$ (\cref{lem:static-retrieval}) that maps each static key to its $b$-bit value, using $(n - s) b + O(\log n + \log \log U)$ bits.
    \item \emph{The values of the dynamic keys.} A function $h : [U] \to [s]$ that is injective on $S_{\mathrm{dyn}}$ (\cref{lem:mphf}), using $s \log e + O(\log n + \log \log U)$ bits, along with an array $A$ of $s$ entries, each of $b$ bits, where $A[h(x)]$ stores the current value of each dynamic key $x$.
\end{itemize}
The memory of the data structure is the concatenation of these components. The length of each component is determined by the parameters $n$, $U$, $b$, and $k$, and by $|S_{\mathrm{dyn}} \cup P|$ (which is stored explicitly), so the components can be parsed.

\paragraph{Operations.}
To answer $\textsc{Query}(x)$, the data structure checks whether $x$ passes the filter and, if so, whether $R_{\mathrm{type}}$ indicates that $x$ is dynamic. If both checks succeed, it returns $A[h(x)]$; otherwise, it returns $R_{\mathrm{stat}}(x)$. To perform $\textsc{Update}(x, v)$ on a dynamic key $x$, the data structure sets $A[h(x)]$ to $v$.

\paragraph{Correctness.}
Since $\textsc{Update}$ modifies only the array $A$, every other component remains as it was at construction time. Consider a query to some key $x \in S$. If $x \in S_{\mathrm{dyn}}$, then $x$ passes the filter, and $R_{\mathrm{type}}$ indicates that $x$ is dynamic, so the query returns $A[h(x)]$. Since $h$ is injective on $S_{\mathrm{dyn}}$, the entry $A[h(x)]$ is modified only by updates to $x$, so it stores the value most recently assigned to $x$. If, on the other hand, $x \in S_{\mathrm{stat}}$, then either $x$ does not pass the filter, or $x \in P$, in which case $R_{\mathrm{type}}$ indicates that $x$ is static. Either way, the query returns $R_{\mathrm{stat}}(x)$, which is the value of $x$ (since static keys are never updated).

\paragraph{Space.}
Summing the sizes of the components, the total space is
$$(n - s) b + sb + s \log k + s \log e + O(s + \log n + \log \log U) = nb + s \log k + O(s + \log n + \log \log U)$$
bits. Since $s = n/k$, this completes the proof of \cref{prop:partially-dynamic-upper}.

Note that, when $U \le 2^{\poly(n)}$, the dominant term in the redundancy, namely $s \log k$, comes from the filter, which is precisely the component that (approximately) encodes which keys are dynamic. This matches the intuition from the lower bound, where the redundancy comes from the $\log \binom{n}{s} = \Theta(s \log k)$ bits needed to encode the set of dynamic keys.

\subsection{Time-Space Tradeoffs}
\label{sec:partially-dynamic-tradeoff}

We now consider the query time. Hu et al.\ \cite{hu2025static} established the following tight bounds on the space of static retrieval as a function of the query time (see Theorems 2.1 and 4.1 of \cite{hu2025static}).

\begin{theorem}[\cite{hu2025static}]
    \label{thm:hu-static}
    Let $v = O(\log n)$ be a positive integer, let $t \ge 1$, and suppose that $U \le n^{O(1)}$.
    \begin{itemize}
        \item \emph{Lower bound.} Suppose that $U \ge 2n$. Any static retrieval data structure in the cell-probe model with word size $w \ge v$ that stores $v$-bit values for $n$ keys from $[U]$, and that answers queries using $t$ probes in expectation (where the data structure may use an arbitrarily long random tape), uses at least $nv + \lfloor n \cdot e^{-O(wt/v)} \rfloor$ bits of space.
        \item \emph{Upper bound.} There is a static retrieval data structure in the word RAM model with word size $w = \Theta(\log n)$ that stores $v$-bit values for $n$ keys from $[U]$, answers queries in $O(t)$ time in the worst case, and uses $nv + O(n / 2^{\Omega(wt/v)} + \log^2 n)$ bits of space. The data structure assumes free access to random hash functions, and its construction succeeds with high probability.
    \end{itemize}
\end{theorem}

Combining \cref{thm:hu-static} with \cref{thm:partially-dynamic} gives tight bounds for partially value-dynamic retrieval.

\begin{theorem}
    \label{thm:partially-dynamic-tradeoff}
    Let $c$ and $k_0$ be the constants from \cref{thm:partially-dynamic}, and let $k \ge \max(k_0, 2)$. Let $b = O(\log n)$, let $t \ge 1$, suppose that $2n \le U \le n^{O(1)}$, and consider a word size $w = \Theta(\log n)$ with $w \ge b$.
    \begin{itemize}
        \item \emph{Lower bound.} Any partially value-dynamic retrieval data structure in the cell-probe model that stores $b$-bit values for $n$ keys from $[U]$, of which $n/k$ are dynamic, and that answers queries using $t$ probes in expectation, uses at least
        $$nb + \max\left\{c \cdot \frac{n}{k} \log k,\ \left\lfloor n \cdot e^{-O(wt/b)} \right\rfloor\right\}$$
        bits of space.
        \item \emph{Upper bound.} There is a partially value-dynamic retrieval data structure in the word RAM model that stores $b$-bit values for $n$ keys from $[U]$, of which $n/k$ are dynamic, answers queries in $O(t)$ time and performs updates in $O(1)$ time in the worst case, and uses
        $$nb + O\left(\frac{n}{k} \log k + \frac{n}{2^{\Omega(wt/b)}} + \log^2 n\right)$$
        bits of space. The data structure assumes free access to random hash functions, and its construction succeeds with high probability.
    \end{itemize}
\end{theorem}

Since the maximum of two quantities is at least their average, the two bounds match, up to the constants in the exponents, constant factors, and the additive $O(\log^2 n)$ term. In other words, the optimal redundancy is the sum of the cost of remembering which keys are dynamic, which is $\Theta((n/k) \log k)$, and the cost of fast queries for static retrieval, which is $n \cdot e^{-\Theta(wt/b)}$.

\begin{proof}[Proof of the lower bound]
    The first term in the maximum is \cref{thm:partially-dynamic}, which applies regardless of the query time. For the second term, observe that a partially value-dynamic retrieval data structure on which no \textsc{Update} operations are performed is, in particular, a static retrieval data structure for the $n$ keys (where, for concreteness, we designate the $n/k$ smallest keys as dynamic). Its queries use $t$ probes in expectation, so \cref{thm:hu-static} implies that it uses at least $nb + \lfloor n \cdot e^{-O(wt/b)} \rfloor$ bits of space.
\end{proof}

\begin{proof}[Proof of the upper bound]
    We consider two cases, depending on whether $k \le \sqrt{n}$. In both cases, we store the values of the static keys in a static retrieval data structure $R_{\mathrm{stat}}$ given by \cref{thm:hu-static}, using $(n - s) b + O(n / 2^{\Omega(wt/b)} + \log^2 n)$ bits, with queries taking $O(t)$ time. (Here, we use that $n - s \ge n/2$, so that $U \le (n - s)^{O(1)}$ and $w = \Theta(\log (n - s))$.)

    \paragraph{Case 1: $k \le \sqrt{n}$.}
    We implement the components of the data structure from \cref{sec:partially-dynamic-upper-space} as follows. Since $s \ge \sqrt{n}$, we have $U \le s^{O(1)}$ and $w = \Theta(\log s)$, so \cref{thm:hu-static} applies to sets of between $s$ and $n$ keys.
    \begin{itemize}
        \item For the filter, we take $g : [U] \to \mathbb{F}_2^\ell$ to be a random hash function, where $\ell = \lceil \log k \rceil \le w$, and implement $R_{\mathrm{fp}}$ using \cref{thm:hu-static} with $v = \ell$ and with $O(1)$ query time. This uses $s\ell + O(s + \log^2 n) \le s \log k + O(s + \log^2 n)$ bits. As in \cref{sec:partially-dynamic-upper-space}, for each static key $y$, the fingerprint $g(y)$ is independent of $R_{\mathrm{fp}}(y)$, so a random choice of $g$ satisfies $|P| \le 2s$ with probability at least $1/2$. The construction tries $\lceil \log n \rceil$ independent random hash functions $g$, and stores the index of the first one that satisfies $|P| \le 2s$, using $O(\log \log n)$ bits. The probability that all of them fail is at most $1/n$.
        \item We implement $R_{\mathrm{type}}$ using \cref{thm:hu-static} with $v = 1$ and with $O(1)$ query time. This uses $|S_{\mathrm{dyn}} \cup P| + O(s + \log^2 n) = O(s + \log^2 n)$ bits.
        \item We implement $h$ using the minimal perfect hash function of Hagerup and Tholey \cite{HT01}, which can be evaluated in $O(1)$ time and uses $O(s + \log \log U) = O(s + \log n)$ bits. Since $b \le w$, each entry of $A$ can be read or written in $O(1)$ time.
    \end{itemize}
    Queries take $O(t)$ time (dominated by the query to $R_{\mathrm{stat}}$), and updates take $O(1)$ time. The total space is $nb + s \log k + O(s + n/2^{\Omega(wt/b)} + \log^2 n)$ bits.

    \paragraph{Case 2: $k > \sqrt{n}$.}
    In this case, $\log n < 2 \log k$, so we can afford to store the dynamic keys explicitly. Fredman, Koml\'os, and Szemer\'edi \cite{fredman1984storing} show how to store a set of $s$ keys from $[U]$ in a table of at most $6s$ cells, each of $O(\log U) = O(\log n)$ bits, such that a membership query to any $x \in [U]$ takes $O(1)$ probes and, if $x$ is in the set, returns the cell in which $x$ is stored. We store $S_{\mathrm{dyn}}$ in this way, along with an array that maps each cell containing a dynamic key to a distinct index in $[s]$, and an array $A$ of $s$ entries, each of $b$ bits, that stores the value of the dynamic key with each index. To answer $\textsc{Query}(x)$, the data structure looks up $x$ in the table. If $x$ is found, then it returns the corresponding entry of $A$; otherwise, it returns $R_{\mathrm{stat}}(x)$. Updates to a dynamic key $x$ modify the corresponding entry of $A$. Queries take $O(t)$ time, updates take $O(1)$ time, and the total space is
    $$(n - s) b + sb + O(s \log n) + O\left(\frac{n}{2^{\Omega(wt/b)}} + \log^2 n\right) = nb + O\left(s \log k + \frac{n}{2^{\Omega(wt/b)}} + \log^2 n\right)$$
    bits.

    In both cases, since $s = n/k$, the space bound matches the theorem statement.
\end{proof}

\section{Assumptions on the Computational Model}
\label{sec:model-wlog}

In this appendix, we justify the assumptions about the computational model that we make in \cref{sec:prelim}.

\paragraph{Fixed-size memory and illegal operations.}
Our lower bounds often reason about what happens when \textsc{Update} is invoked on a key that the user is not permitted to update. To make this meaningful, we assume that the memory always consists of exactly $M$ bits, and that \textsc{Update} and \textsc{Query} are deterministic functions defined on \emph{all} inputs, including illegal ones, so that any sequence of operations, legal or not, takes a memory configuration to another memory configuration of exactly $M$ bits. Any data structure can be converted into one with these properties, at the cost of $O(1)$ additional bits, without changing its behavior on legal operation sequences: pad the memory to a fixed length, and have any \textsc{Update} that would produce a memory configuration of more than $M$ bits instead leave the memory unchanged.

\paragraph{Randomization.}
We assume that the data structure is deterministic. This is without loss of generality: if the data structure is randomized, then the communication protocols in this paper can simply treat its random bits as public randomness shared between Alice and Bob. The same is true even if the data structure's preprocessing is permitted to fail with probability up to $1/2$ (meaning that, with probability at most $1/2$, the resulting data structure answers some sequence of legal operations incorrectly). In this case, Alice can detect the failure and fall back to sending her input explicitly. Since the explicit encoding of Alice's input is within $O(1)$ bits of optimal, this changes our lower bounds by at most $O(1)$ bits.

\end{document}